\documentclass[english]{IEEEtran}
\usepackage[T1]{fontenc}
\usepackage[latin9]{inputenc}
\usepackage{amsthm}
\usepackage{amsmath}
\usepackage{amssymb}
\usepackage{graphicx}

\makeatletter
\theoremstyle{plain}
\newtheorem{thm}{\protect\theoremname}
\theoremstyle{plain}
\newtheorem{cor}[thm]{\protect\corollaryname}
\theoremstyle{plain}
\newtheorem{lem}[thm]{\protect\lemmaname}
\theoremstyle{remark}
\newtheorem{rem}[thm]{\protect\remarkname}

\usepackage{cite}
\allowdisplaybreaks

\makeatother

\usepackage{babel}
\providecommand{\corollaryname}{Corollary}
\providecommand{\lemmaname}{Lemma}
\providecommand{\remarkname}{Remark}
\providecommand{\theoremname}{Theorem}

\begin{document}

\title{Transport Capacity of Distributed Wireless CSMA Networks}

\author{Tao Yang, Guoqiang Mao, \emph{Senior Member, IEEE}, Wei Zhang, \emph{Senior
Member IEEE} and Xiaofeng Tao, \emph{Senior Member IEEE}%
\thanks{T. Yang is with the School of Electrical and Information Engineering,
The University of Sydney. Email: tao.yang@sydney.edu.au.%
}\emph{}%
\thanks{G. Mao is with the School of Computing and Communications and Center
for Real-time Information Networks, The University of Technology,
Sydney, and National ICT Australia. Email: guoqiang.mao@uts.edu.au.%
}\emph{}%
\thanks{W. Zhang is with the School of Electrical Engineering \& Telecommunications,
The University of New South Wales. Email: wzhang@ee.unsw.edu.au.%
}\emph{}%
\thanks{X. Tao is with National Engineering Lab. for Mobile Network Security, Beijing University of Posts and Telecommunications. Email: taoxf@bupt.edu.cn.%
}\emph{}%
\thanks{This research is funded by ARC Discovery projects: DP110100538 and
DP120102030.%
}}
\maketitle
\begin{abstract}
In this paper, we study the transport capacity of large multi-hop
wireless CSMA networks. Different from previous studies which rely
on the use of centralized scheduling algorithm and/or centralized
routing algorithm to achieve the optimal capacity scaling law, we
show that the optimal capacity scaling law can be achieved using entirely
distributed routing and scheduling algorithms. Specifically, we consider
a network with nodes Poissonly distributed with unit intensity on
a $\sqrt{n}\times\sqrt{n}$ square $B_{n}\subset\Re^{2}$. Furthermore,
each node chooses its destination randomly and independently and transmits
following a CSMA protocol. By resorting to the percolation theory
and by carefully tuning the three controllable parameters in CSMA
protocols, i.e. transmission power, carrier-sensing threshold and
count-down timer, we show that a throughput of $\Theta\left(\frac{1}{\sqrt{n}}\right)$
is achievable in distributed CSMA networks. Furthermore, we derive
the pre-constant preceding the order of the transport capacity by
giving an upper and a lower bound of the transport capacity. The tightness
of the bounds is validated using simulations.\end{abstract}

\section{Introduction\label{sec:Introduction}}

Wireless multi-hop networks have been increasingly used in civilian
and military applications. In a wireless multi-hop network, nodes
communicate with each other via wireless multi-hop paths, and packets
are forwarded collaboratively hop-by-hop by intermediate relay nodes
from sources to their respective destinations. Studying the capacity
of these networks is an important problem. 

Capacity of large wireless networks has been extensively investigated
with a particular focus on the throughput scaling laws when the network
becomes sufficiently large \cite{Chau11Capacity,Cho06Capacity,Dousse06On,Franceschetti07Closing,Grossglauser02Mobility,Gupta00Capacity,Gupta00Internets,Hu10A,Kulkarni04A,Li09Impacts,Xie04A}.
Two metrics are widely used in the study of network capacity: transport
capacity and transmission capacity. The transport capacity quantifies
the end-to-end throughput that can be achieved between source-destination
pairs whereas the transmission capacity, often used together with
another metric\emph{ }outage probability, quantifies the achievable
single-hop rates in large wireless networks. The transport capacity
is useful to capture the impact of network topology, routing and scheduling
algorithms on network capacity \cite{Alfano09Capacity,Chau11Capacity,Cho06Capacity,Dousse06On,Franceschetti07Closing,Grossglauser02Mobility,Gupta00Capacity,Hu10A,Kulkarni04A,Li09Impacts,Li11The,Yang12Capacity}.
Comparatively, the transmission capacity is more useful when the focus
is on the impact of physical layer details, e.g., fading, interference
and signal propagation model, on the capacity of large networks \cite{Andrews10Random,Ganti11High,Ganti09Interference,Weber10An,Baccelli09Stochastic}.
In this paper, we focus on the study of the transport capacity.

In the ground-breaking work \cite{Gupta00Capacity} by Gupta and Kumar,
it was shown that in a static network of $n$ nodes uniformly and
i.i.d. on an area of unit size and each node is capable of transmitting
at $W$ bits/second and using a fixed and identical transmission range,
the achievable per-node throughput is $\Theta\left(\frac{W}{\sqrt{n\log n}}\right)$
when each node chooses its destination randomly and independently.
If nodes are optimally and deterministically placed to maximize capacity,
the achievable per-node throughput becomes $\Theta\left(\frac{W}{\sqrt{n}}\right)$.
In a more general setting, assuming only that power attenuates with
distance following a power-law relationship, Xie and Kumar \cite{Xie04A}
showed that $\Theta\left(\frac{1}{\sqrt{n}}\right)$ is an upper bound
on the per-node throughput of wireless networks, regardless of the
scheduling and routing algorithm being employed. Since then, a number
of solutions have been proposed to achieve the above upper bounds
under various network settings and using various routing and scheduling
algorithms \cite{Alfano09Capacity,Chau11Capacity,Cho06Capacity,Dousse06On,Franceschetti07Closing,Grossglauser02Mobility,Gupta00Capacity,Hu10A,Kulkarni04A,Li09Impacts,Li11The,Yang12Capacity}.
In \cite{Franceschetti07Closing}, Franceschetti \emph{et al.} considered
the same network as that in \cite{Gupta00Capacity} except that nodes
are allowed to use two different transmission ranges. They showed
that by using a routing scheme based on the so-called ``highway system''
and a centralized/deterministic time division multiple access (TDMA)
scheme, the per-node throughput can reach $\Theta\left(\frac{1}{\sqrt{n}}\right)$
even when nodes are randomly located. Specifically, the highway system
is formed by nodes using the smaller transmission range, whereas the
larger transmission range is used for the last mile, i.e., between
the source (or destination) and its nearest highway node. The existence
of highway system was established using the percolation theory. 

Other work in the field includes \cite{Grossglauser02Mobility} in
which Grossglauser and Tse showed that in mobile networks, by leveraging
on the nodes' mobility, a per-node throughput of $\Theta\left(1\right)$
can be achieved at the expense of large delay. Their work \cite{Grossglauser02Mobility}
has sparked huge interest in studying the capacity-delay tradeoffs
in mobile networks assuming various mobility models and the obtained
results often vary greatly with the different mobility models being
considered, see \cite{Jacquet12On,Kong08Connectivity,Li09Capacity,Neely05Capacity,EI06Optimal,EI06OptimalThroughput}
and references therein for examples. In \cite{Chen09Order}, Chen
\emph{et al.} studied the capacity of wireless networks under a different
traffic distribution. In particular, they considered a set of $n$
randomly deployed nodes transmitting to a single sink or multiple
sinks where the sinks can be either regularly deployed or randomly
deployed. They showed that with single sink, the transport capacity
is given by $\Theta\left(W\right)$; with $k$ sinks, the transport
capacity is increased to $\Theta\left(kW\right)$ when $k=O(n\log n)$
or $\Theta\left(n\log nW\right)$ when $k=\Omega\left(n\log n\right)$.
Furthermore, there is also significant amount of work studying the
impact of infrastructure nodes \cite{Zemlianov05Capacity} and multiple-access
protocols \cite{Durvy09On,Alfano11New} on the capacity and the multicast
capacity \cite{Li09Multicast}. We refer readers to \cite{Haenggi09Stochastic}
for a comprehensive review of related work.

The above work of Franceschetti \emph{et al.} \cite{Franceschetti07Closing}
and Gupta and Kumar \cite{Gupta00Capacity,Xie04A}, and most other
work in the field \cite{Alfano09Capacity,Cho06Capacity,Dousse06On,Hu10A,Kulkarni04A,Li09Impacts,Li11The},
established the capacity of wireless multi-hop networks using centralized
scheduling and routing schemes, which may not be appropriate for large-scale
networks being investigated in \cite{Franceschetti07Closing,Gupta00Capacity,Xie04A}. 

In a recent work \cite{Chau11Capacity}, Chau \emph{et al.} took the
lead in studying the throughput of CSMA networks. They showed that
CSMA networks can achieve the per-node throughput $\Theta\left(\frac{1}{\sqrt{n}}\right)$,
the same order as networks using optimal centralized TDMA, if multiple
back off countdown rates are used in the distributed CSMA protocol
and packets are routed using the highway system proposed in \cite{Franceschetti07Closing}.
While the use of distributed CSMA for scheduling in \cite{Chau11Capacity}
constitutes a significant advance compared with the centralized TDMA
considered in previous work, the routing scheme in \cite{Chau11Capacity}
still relies on the highway system, which needs centralized coordination
to identify the highway nodes and to establish the highway. The centralized
routing scheme used in \cite{Chau11Capacity} is not compatible with
the distributed CSMA scheduling scheme. In this sense, the routing
and scheduling scheme in \cite{Chau11Capacity} is not entirely distributed
and may not be suitable for large-scale networks. Furthermore, the
deployment of the highway system in CSMA networks requires two different
carrier-sensing ranges to be used: a smaller carrier-sensing range
used by the highway nodes and a larger carrier-sensing range used
by the remaining nodes to access the highway. The use of two different
carrier-sensing ranges may exacerbate the hidden node problem in CSMA
networks, which will be explained in detail in Section \ref{sec:Hidden-node-free-design}.
To conquer the potential hidden problem brought by the use of two
different carrier-sensing ranges, the entire frequency bandwidth is
divided into two sub-bands for use by the two types of nodes employing
different carrier-sensing ranges respectively. This imposes additional
hardware requirements on the nodes and also causes spectrum waste.

Based on the above observation, we are motivated to develop a distributed
scheduling and routing algorithm to achieve the order-optimal throughput
in CSMA networks in this paper. Specifically, by resorting to the
percolation theory and by carefully tuning the three controllable
parameters in CSMA protocols, i.e., transmission power, carrier-sensing
threshold and count-down timer, we show that a throughput of $\Theta\left(\frac{1}{\sqrt{n}}\right)$
is achievable in distributed CSMA networks. Furthermore, we analyze
the pre-constant preceding the order of the transport capacity by
giving an upper and a lower bound of the transport capacity. The tightness
of the bounds is established using simulations.

The following is a detailed summary of our contributions:
\begin{itemize}
\item We develop a distributed routing and scheduling algorithm that is
able to achieve the order-optimal throughput in CSMA networks. More
specifically, the routing decision relies on the use of local neighborhood
knowledge only and each node competes for channel access in a distributed
and randomized manner using CSMA protocols.
\item We demonstrate that by jointly tuning the carrier-sensing threshold
and the transmission power, the hidden node problem can be eliminated
even for nodes using different carrier-sensing thresholds, different
transmission powers and a common frequency band. This is different
from the techniques used in the previous work \cite{Chau11Capacity}
where nodes using different carrier-sensing ranges have to use different
frequency band for transmission. The technique developed provides
guidance on setting the carrier-sensing threshold and the transmission
power to avoid the hidden node problem in CSMA networks in a more
general setting.
\item We analyze the pre-constant preceding the order of the transport capacity
by giving an upper and a lower bound of the transport capacity. As
pointed out in \cite{Haenggi09Stochastic}, the pre-constant is important
to fully understand the impact of various parameters on network capacity.
\item Extensive simulations are carried out which validate the tightness
of our analytical results.
\end{itemize}
The rest of this paper is organized as follows. Section \ref{sec:Related-work}
reviews related work; Section \ref{sec:Network-Model} presents the
network model and defines notations and concepts used in the later
analysis; Section \ref{sec:Routing Scheme} describes the routing
algorithm and analyzes the traffic load of each node; Section \ref{sec:Hidden-node-free-design}
presents the solution for obtaining a hidden node free CSMA network;
Section \ref{sec:Capacity-Analysis} optimizes the medium access probability
for each node by tuning the back off timer and analyzes the per-node
throughput under our proposed communication strategy; Finally, Section
\ref{sec:Conclusion} concludes the paper.

\section{Related Work\label{sec:Related-work}}

In addition to the work mentioned in Section \ref{sec:Introduction}
on general studies of network capacity, in this section we further
review work closely related to research and theoretical analysis in
this paper.

Limited work exists on analyzing capacity of large networks running
distributed routing and scheduling algorithms, despite their extensive
deployment in real networks. Reference \cite{Chau11Capacity} discussed
in Section \ref{sec:Introduction} was among the first work studying
the capacity of networks employing distributed and randomized CSMA
protocols and showed that these networks can achieve the same order-optimal
throughput of $\Theta\left(\frac{1}{\sqrt{n}}\right)$ as networks
employing centralized TDMA schemes. Yang \emph{et al} \cite{Yang12Capacity}
studied the achievable throughput of three dimensional CSMA networks.
Ko \emph{et al} \cite{Ko13Optimization} showed that in CSMA networks,
by jointly optimizing the transmission range and packet generation
rate, the end-to-end throughput and end-to-end delay can scale as
$\Theta\left(\frac{1}{\sqrt{n\log n}}\right)$ and $\Theta\left(\frac{n}{\sqrt{\log n}}\right)$,
respectively. Byun \emph{et al} \cite{Byun13Delay} showed that distributed
slotted ALOHA protocols can have order-optimal throughput. Unlike
in ALOHA, where each node access the medium \emph{independently} with
a prescribed probability, nodes of CSMA networks suffer from a\emph{
spatial correlation problem}, which means that the activity of a node
is dependent on the activities of other nodes due to the carrier-sensing
operation. This correlation problem makes the analysis of interference
and capacity of CSMA networks more challenging than that of ALOHA
networks. Therefore, although both ALOHA and CSMA are distributed
medium access control protocols, the results obtained for ALOHA networks
are not directly applicable to CSMA networks.

Some research efforts were also devoted to modeling the spatial distribution
of concurrent transmitters obeying carrier-sensing constraints and
the distribution of interference resulting from these transmitters.
The spatial distribution of concurrent transmitters following CSMA
protocols are often modeled by the Mat�rn hard-core point process
(p.p.) and sometimes \emph{approximated} by the Poisson point process
\cite{Nguyen12On,Busson09Point,Baccelli09Stochastic,Haenggi09Interference,Haenggi11Mean}.
In more recent studies \cite{Busson09Point,Alfano11New,Haenggi11Mean},
the Random Sequential Absorption (RSA) p.p. was proposed as a more
natural model for representing the spatial distribution of concurrent
CSMA transmitters. Nguyen and Baccelli \cite{Nguyen12On} studied
the RSA p.p. by characterizing its generating functional and derived
upper and lower bounds for the generating functional. Furthermore,
the authors of \cite{Nguyen12On} derived the network performance
metrics, viz., average medium access probability and average transmission
success probability (two commonly used metrics in the study of transmission
capacity), in terms of the generating functional. Alfano \emph{et
al.} in \cite{Alfano11New} obtained \emph{approximately} the transmission
capacity distribution. The above work \cite{Nguyen12On,Alfano11New}
studied the transmission capacity by investigating the transmission
success probability and the medium access probability of a \emph{typical}
node, which quantifies the spatial \emph{average performance} of the
network. In comparison, the transport capacity often quantifies the
throughput that can be achieved by \emph{every} source-destination
pair (asymptotically almost surely), which is often associated with
the \emph{worst case} performance.

Improving spatial frequency reuse of CSMA networks is an important
problem that has also been extensively investigated, see \cite{Alawieh09Improving,Kim08Understanding,Lin07Interplay}
for the relevant work. However, high level of spatial frequency reuse
does not directly lead to increased end-to-end throughput because
the latter performance metric also critically relies on the communication
strategies, i.e., routing algorithm and scheduling scheme, used in
the network. In this paper we focus on the study of achievable end-to-end
throughput.

\section{Network Model and Settings\label{sec:Network-Model}}

In this section, we introduce the network model, the signal propagation
model, the SINR model and define notations and concepts that are used
in later analysis.

Two network models are widely used in the study of (asymptotic) network
capacity: the \emph{dense network model} and the \emph{extended network
model}. By appropriate scaling of the distance, the results obtained
under one model can often be extended to the other one \cite{Franceschetti07Random}.
In this paper, we consider the extended network model. Particularly
we consider a network with nodes deployed on a $\sqrt{n}\times\sqrt{n}$
box $B_{n}\subset\Re^{2}$ according to a Poisson point process with
unit intensity. Each node chooses its destination randomly and independently
of other nodes. We study the capacity of the above network as $n\rightarrow\infty$.
It is assumed that all data transmissions are conducted over a common
wireless channel.

We are mainly concerned with the events that occur inside $B_{n}$
asymptotically almost surely (a.a.s.) as $n\rightarrow\infty$. An
event $\xi_{n}$ depending on $n$ is said to occur a.a.s. if and
only if (iff) its probability approaches $1$ as $n\rightarrow\infty$.
The following notations are used throughout the paper concerning the
asymptotic behavior of positive functions:
\begin{itemize}
\item $f\left(n\right)=O\left(g\left(n\right)\right)$ if that there exist
a positive constant $c$ and an integer $n_{0}$ such that $f\left(n\right)\leq cg\left(n\right)$
for any $n>n_{0}$;
\item $f\left(n\right)=\Omega\left(g\left(n\right)\right)$ if $g\left(n\right)=O\left(f\left(n\right)\right)$;
\item $f\left(n\right)=\Theta\left(g\left(n\right)\right)$ if that there
exist two constants $c_{1}$, $c_{2}$ and an integer $n_{0}$ such
that $c_{1}g\left(n\right)\leq f\left(n\right)\leq c_{2}g\left(n\right)$
for any $n>n_{0}$;
\item $f\left(n\right)=o\left(g\left(n\right)\right)$ if ${\displaystyle \lim_{n\rightarrow\infty}}\frac{f\left(n\right)}{g\left(n\right)}=0$.
\end{itemize}

\subsection{Interference model\label{sub:Interference-model}}

Let $\boldsymbol{x}_{k},\, k\in\Gamma$, be the location of node $k$,
where $\Gamma$ represents the set of indices of all nodes. When node
$i$ is transmitting with power $P_{i}$, the received power at node
$j$ located at $\boldsymbol{x}_{j}$ from node $i$ is given by $P_{i}\left\Vert \boldsymbol{x}_{i}-\boldsymbol{x}_{j}\right\Vert ^{-\alpha}$
where $\left\Vert \boldsymbol{x}_{i}-\boldsymbol{x}_{j}\right\Vert ^{-\alpha}$
represents the path-loss from node $i$ to node $j$, $\alpha$ is
the path-loss exponent and $\left\Vert \boldsymbol{x}_{i}-\boldsymbol{x}_{j}\right\Vert $
is the Euclidean distance between the two nodes. In the paper we assume
that $\alpha>2$. This channel model is widely used in the literature
\cite{Haenggi09Interference,Franceschetti07Closing,Gupta00Capacity,Chau11Capacity}.
A transmission from node $i$ to node $j$ is successful iff the SINR
at node $j$ is above a predetermined threshold $\beta$, i.e.,
\begin{equation}
\textrm{SINR}\left(\boldsymbol{x}_{i}\rightarrow\boldsymbol{x}_{j}\right)=\frac{P_{i}\left\Vert \boldsymbol{x}_{i}-\boldsymbol{x}_{j}\right\Vert ^{-\alpha}}{N_{0}+\underset{k\in\mathcal{T}_{i}}{\sum}P_{k}\left\Vert \boldsymbol{x}_{k}-\boldsymbol{x}_{j}\right\Vert ^{-\alpha}}\geq\beta\label{eq:SINR}
\end{equation}
where $\mathcal{T}_{i}\subseteq\Gamma$ denotes the set of simultaneously
active transmitters as node $i$ and $N_{0}$ represents background
noise. In this paper we consider an interference-limited network and
assume that the impact of $N_{0}$ is negligible. Despite the common
knowledge that a higher SINR can lead to an increased link capacity,
in reality transmission from a transmitter to a receiver can only
occur at one of a set of preset data rates after the SINR threshold
is met \cite{Kim08Understanding,Lin07Interplay}. Therefore for a
transmitter-receiver pair, when its associated SINR is above $\beta$,
it is considered that the transmitter can transmit to the receiver
at a fixed rate of 
\begin{equation}
W=\log_{2}\left(1+\beta\right)\;\; b/s\label{eq:data rate definition}
\end{equation}

\subsection{Definition of throughput\label{sub:Definition-of-throughput}}

Each node sends packets to an independently and randomly chosen destination
node via multiple hops. A node can be a source node, a destination
node for another source node, a relay node or a mixture. 
\begin{quotation}
The \emph{per-node} \emph{throughput} or equivalently the \emph{transport
capacity} of the network, denoted by $\lambda\left(n\right)$, is
defined as the \emph{maximum} rate that could be achieved a.a.s. by
\emph{all} source-destination pairs simultaneously. Similar as that
in \cite{Gupta00Capacity}, we say that a per-node throughput of $\lambda\left(n\right)$
is \emph{feasible} if there is a temporal and spatial routing and
scheduling scheme such that every node can send $\lambda\left(n\right)$
bits/sec \emph{on time average} to its destination a.a.s., i.e., there
exists a sufficiently large positive number $\tau$ such that in every
finite time interval $\left[\left(j-1\right)\tau,j\tau\right]$ every
node can send $\tau\lambda\left(n\right)$ bits to its destination
a.a.s.. 
\end{quotation}

\subsection{CSMA protocol}

The general idea of CSMA protocols is that before transmission, a
node will sense other active transmissions in its vicinity such that
nearby nodes will not transmit simultaneously. More specifically,
a node $j$ is said to be \emph{in contention with} node $i$ if the
received power by node $i$ from node $j$ is above the carrier-sensing
threshold $T_{i}$ of node $i$, i.e., 
\begin{equation}
P_{j}\left\Vert \boldsymbol{x}_{i}-\boldsymbol{x}_{j}\right\Vert ^{-\alpha}>T_{i}.\label{eq:carrier-sensing threshold}
\end{equation}
Node $i$ can only transmit if it senses no other active transmissions
in contention, or in other words the node senses the channel idle. 

To prevent the situation where several nearby nodes simultaneously
start transmitting when their common neighbor stops its transmission,
hence causing a collision, a \emph{back off} mechanism is often employed
such that a node sensing the channel idle will wait a random amount
of time before starting its transmission. 

The following \emph{back off} mechanism is considered in this paper.
Each node senses the channel continuously and maintains a countdown
timer, which is initialized to a non-negative random value. The timer
of a node counts down when it senses the channel idle; when the channel
is sensed as busy, the node freezes its timer. A node initiates its
transmission when its countdown timer reaches zero \emph{and} the
channel is sensed as idle. After finishing its transmission, the node
resets its countdown timer to a new random value for the next transmission.
The distribution of the random initial countdown timer will be specified
later in the paper.

\section{Routing Algorithm and Traffic load\label{sec:Routing Scheme}}

In this section we describe the routing algorithm to be used and analyze
the traffic load for each node under the algorithm. The routing algorithm
chooses the sequence of nodes to deliver a packet from its source
to its destination without considering physical layer implementation
details.

\begin{figure}
\begin{centering}
\includegraphics[width=0.9\columnwidth]{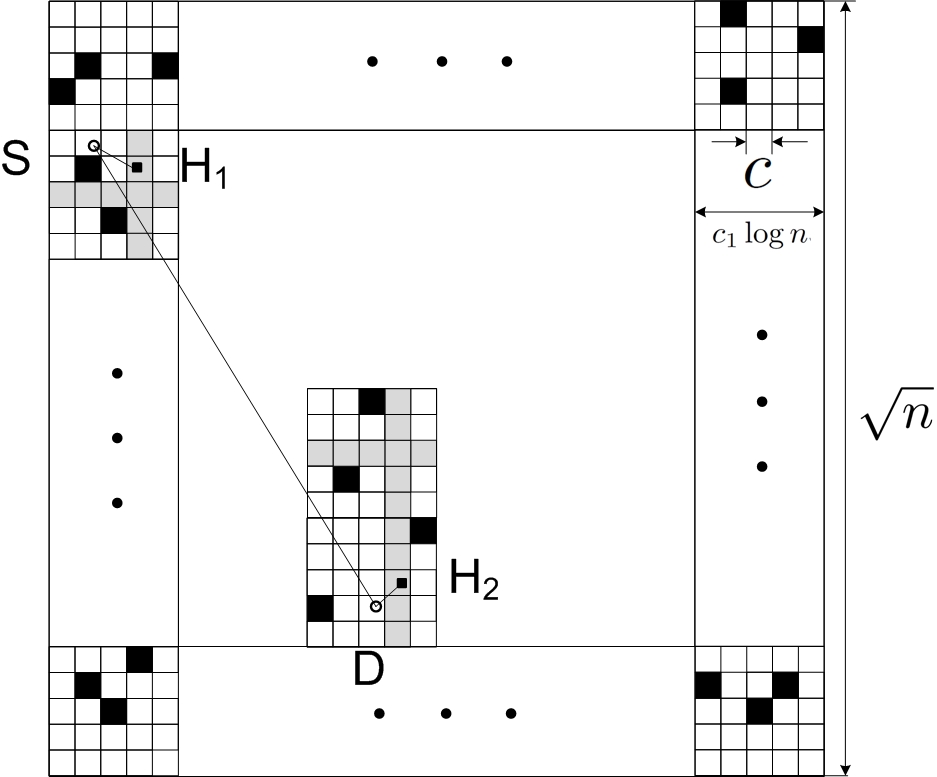}
\par\end{centering}

\protect\caption{\label{fig:partition}An illustration of partition of $B_{n}$ and
the routing algorithm. Black square represents a closed site and white
square represents an open site. Grey square represents an open site
that forms an open path. $S$ and $D$, indicated by two small hollow
circles, are a pair of source and destination nodes. $H_{1}$ and
$H_{2}$, indicated by two small black squares, are two nodes located
in open sites that form open paths. First $S$ transmits its packets
to $H_{1}$ using a transmission range of up to $\sqrt{2}c_{1}\log n$.
Then the packets will be routed along the open paths to $H_{2}$,
using a transmission range of up to $\sqrt{5}c$. Finally, $H_{2}$
transmits the packets to the destination $D$. If $H_{1}$ itself
is a source node, then it transmits its packet directly to the next-hop
node along the open path, using a transmission range of up to $\sqrt{5}c$. }
\end{figure}

To begin the construction of our routing algorithm, we partition the
box $B_{n}$ of size $\sqrt{n}\times\sqrt{n}$ into \emph{squares}
of side length $c_{1}\log n$ where $c_{1}$ is a positive constant.
Each of these squares is then further subdivided into smaller \emph{cells}
of constant side length $c$. The values of $c_{1}$ and $c$ will
be specified later. See Fig. \ref{fig:partition} for an illustration.
Following common terminology used in the percolation theory, we also
refer to these cells as \emph{sites} and use the two terms cells and
sites exchangeably. We call a site \emph{open} if it contains at least
one node, and \emph{closed} otherwise. Due to the Poisson distribution
of nodes with unit intensity, it can be easily obtained that a site
is open with probability $p\triangleq1-e^{-c^{2}}$. Furthermore,
the event that a site is open or closed is independent of the event
that another distinct site is open or closed. The total number of
sites in a square is $\left(\frac{c_{1}}{c}\log n\right)^{2}$, the
total number of sites in $B_{n}$ is $\left(\frac{\sqrt{n}}{c}\right)^{2}$
and the total number of squares in $B_{n}$ is $\left(\frac{\sqrt{n}}{c_{1}\log n}\right)^{2}$.
The techniques to handle the situation that $\frac{c_{1}}{c}\log n$,
$\frac{\sqrt{n}}{c}$ and $\frac{\sqrt{n}}{c_{1}\log n}$ are not
integers are well-known \cite{Franceschetti07Closing}. Therefore
in this paper we ignore some trivial discussions involving the situations
that $\frac{c_{1}}{c}\log n$, $\frac{\sqrt{n}}{c}$ and $\frac{\sqrt{n}}{c_{1}\log n}$
are not integers and consider them to be integers.

Before we can further explain our routing algorithm, we need to first
establish some preliminary results. The network area $B_{n}$ can
be sliced into horizontal rectangles of size $c_{1}\log n\times\sqrt{n}$,
where each horizontal rectangle consists of $\frac{\sqrt{n}}{c_{1}\log n}$
squares. Denote by $H_{i}$ the $i$-th horizontal rectangle where
$1\leq i\leq\frac{\sqrt{n}}{c_{1}\log n}$. We call two sites \emph{adjacent}
if they share a common edge. We define \emph{a left to right open
path }in $H_{i}$ as a sequence of distinct and adjacent open sites
that starts from an open site on the left border of $H_{i}$ and ends
at an open site on the right border of $H_{i}$. The following theorem,
due to \cite[Theorem 4.3.9]{Franceschetti07Random}, gives a lower
bound on the number of open paths in $H_{i}$.
\begin{thm}
\cite[Theorem 4.3.9]{Franceschetti07Random}\label{lem:number of open paths}Consider
site percolation with parameter $p=1-e^{-c^{2}}$. For $c$ sufficiently
large, there exist constants $c_{1}$ and $\omega_{1}$ independent
of $n$, satisfying 
\begin{equation}
\frac{5}{6}<p<1,\label{eq:condition_1}
\end{equation}
 
\begin{equation}
2+c_{1}\log\left(6\left(1-p\right)\right)<0\label{eq:condition_2}
\end{equation}
 and 
\begin{equation}
\omega_{1}\log\frac{p}{1-p}+c_{1}\log\left(6\left(1-p\right)\right)+2<0,\label{eq:condition_3}
\end{equation}
such that a.a.s. there exist at least $\omega_{1}\log n$ left to
right disjoint open paths in every horizontal rectangle. 
\end{thm}
By symmetry, if we partition $B_{n}$ into $\frac{\sqrt{n}}{c_{1}\log n}$
vertical rectangles. Each one is of size $c_{1}\log n\times\sqrt{n}$
and consists of $\frac{\sqrt{n}}{c_{1}\log n}$ squares. Denote by
$V_{j}$ the $j$-th, $1\leq j\leq\frac{\sqrt{n}}{c_{1}\log n}$ vertical
rectangle. It can also be established that a.a.s. there are at least
$\omega_{1}\log n$ top to bottom disjoint open paths in every $V_{j},\:1\leq j\leq\frac{\sqrt{n}}{c_{1}\log n}$.
The following result can be readily established \cite{Franceschetti07Closing}:
\begin{cor}
\label{cor:number of vertical and horizontal paths}There are a.a.s
at least $\omega_{1}\log n$ left-to-right open paths and $\omega_{1}\log n$
top-to-bottom open paths in every square.
\end{cor}

\subsection{Description of the distributed routing algorithm}

We are now ready to explain our routing algorithm. Denote by $SD_{i}$
the line segment connecting node $i$ to its destination. The packets
generated by source node $i$ are routed along the squares intersecting
$SD_{i}$. A square will only serve the traffic of a source-destination
pair if the associated SD line intersects the square. Note that it
is trivial to establish that a.a.s. every square has at least one
node. 

The routing can be divided into three stages:

In the first stage, a source node S, if it is not a node located in
an open site that forms one of the open paths, will transmit its packet
to a node in a \emph{randomly chosen} open site that forms an open
path. If there are multiple nodes in an open site, a node will be
designated randomly to relay all traffic passing through the site.
If the source node is already in a site that forms an open path, this
stage of routing can be omitted and the routing proceeds directly
to the next stage. The maximum distance between the source node and
its next-hop node in this stage is bounded by $\sqrt{2}c_{1}\log n$
because the distance between any two nodes located in a square is
at most $\sqrt{2}c_{1}\log n$. 

In the second stage, the packet will be routed to the adjacent square
intersecting the SD line along one of these left-to-right open path
or top-to-bottom open paths until the packet reaches a node in the
next square. Depending on the location of the open path containing
the relay node and the location of the adjacent square, the packet
may be routed along a left-to-right open path (when the adjacent square
is on the left or on the right of the current square) or along a top-to-bottom
open path (when the adjacent square is on the top or on the bottom
of the current square). If the packet needs to be switched from a
left-to-right open path to a top-to-bottom open path (e.g., when the
previous square is on the left of the current square but the next
square is on the bottom of the current square), a top-to-bottom open
path is chosen randomly from the at least $\omega_{1}\log n$ open
path available. The above process continues until the packet reaches
the square that contains the destination node. In this stage, the
maximum distance between a node and its next-hop node is bounded by
$\sqrt{5}c$ because the distance between any two nodes located in
two adjacent cells is at most $\sqrt{5}c$. 

In the third stage, after reaching the square containing the destination
node, if the destination node is located on one of the open paths,
the packet will be routed along a multi-hop path to the destination
via open paths; if the destination is not located on one of the open
paths, the packet will be transmitted to the destination directly
and the maximum transmission distance is bounded by $\sqrt{2}c_{1}\log n$. 

The same route is used for all packets belonging to the same source-destination
pair.

The feasibility of the above routing algorithm is guaranteed by Corollary
\ref{cor:number of vertical and horizontal paths}. A node only needs
neighborhood information of nodes no more than $\sqrt{5}c_{1}\log n$
away to make a routing decision. The required information for making
a proper routing decision is vanishingly small compared with that
in the highway algorithm. Furthermore, compared with the network size,
the required information is also vanishingly small as $n\rightarrow\infty$.
Therefore the routing algorithm can be executed in a distributed manner.\textbf{
}On the other hand, we readily acknowledge that neighborhood information
of nodes up to $\sqrt{5}c_{1}\log n$ away\textbf{ }may be required
by the routing algorithm.\textbf{ }The required neighborhood information
grows logarithmically with the size of the network.

\begin{figure}
\begin{centering}
\includegraphics[width=1\columnwidth]{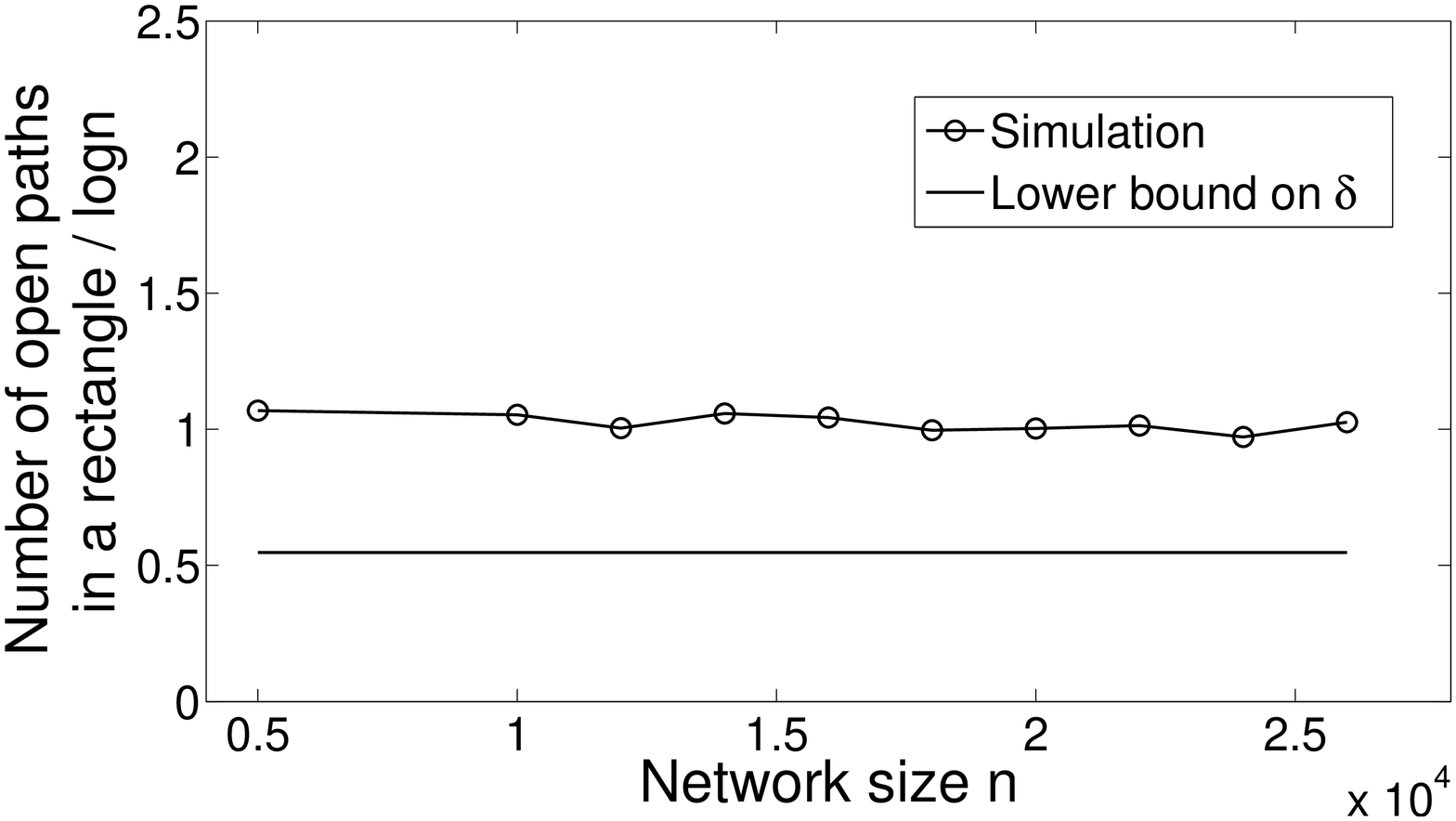}
\par\end{centering}

\protect\caption{\label{fig:number of open paths}An illustration of the number of
left-to-right open paths in a horizontal rectangle as the network
size varies. Vertical axis shows the ratio of the number of open paths
to $\log n$.}
\end{figure}

Corollary \ref{cor:Number-Open-Paths} is a ready consequence of Theorem
\ref{lem:number of open paths}:
\begin{cor}
\label{cor:Number-Open-Paths}Let $c=1.7308$ and $c_{1}=3$, a.a.s.
there are at least $0.5474\log n$ left-to-right open paths in every
horizontal rectangle.
\end{cor}
In the rest of this paper, we carry out analysis assuming that $c$
and $c_{1}$ take values specified in Corollary \ref{cor:Number-Open-Paths}
and $\omega_{1}=0.5474$. Fig. \ref{fig:number of open paths} shows
simulation results of the number of open paths in a horizontal rectangle
as the network size $n$ varies. Each random simulation is repeated
a large number of times and the average result is shown. The confidence
interval is very small and negligible, and thus not plotted in the
figure. The lower bound on the number of open paths suggested in Corollary
\ref{cor:Number-Open-Paths} is also plotted for comparison. As shown
in Fig. \ref{fig:number of open paths}, the lower bound is reasonably
tight. 

\begin{figure}
\begin{centering}
\includegraphics[width=1\columnwidth]{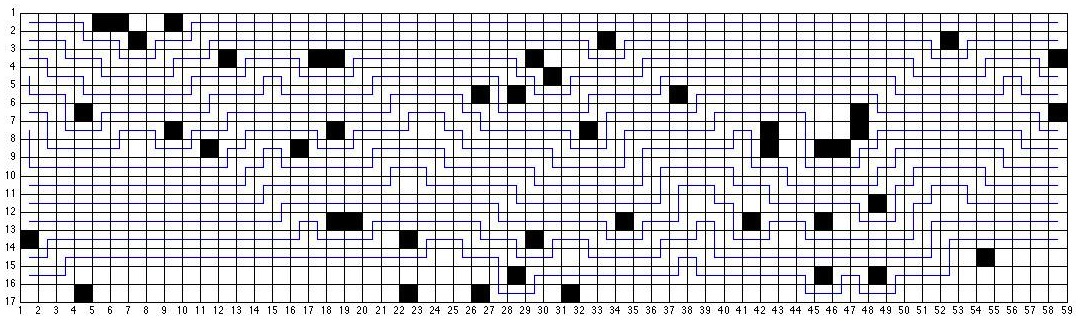}
\par\end{centering}

\protect\caption{\label{fig:Open paths}An illustration of left-to-right open paths
in a rectangle obtained by computer simulations. Black cells represent
closed sites while white cells represent open sites. }

\end{figure}

Fig. \ref{fig:Open paths}, drawn from a simulation, further gives
an intuitive illustration of the open paths in a horizontal rectangle. 

After establishing the routing algorithm, next we analyze the traffic
load for each node under the algorithm, which forms a key step in
analyzing the network capacity.

\begin{figure}
\begin{centering}
\includegraphics[width=1\columnwidth]{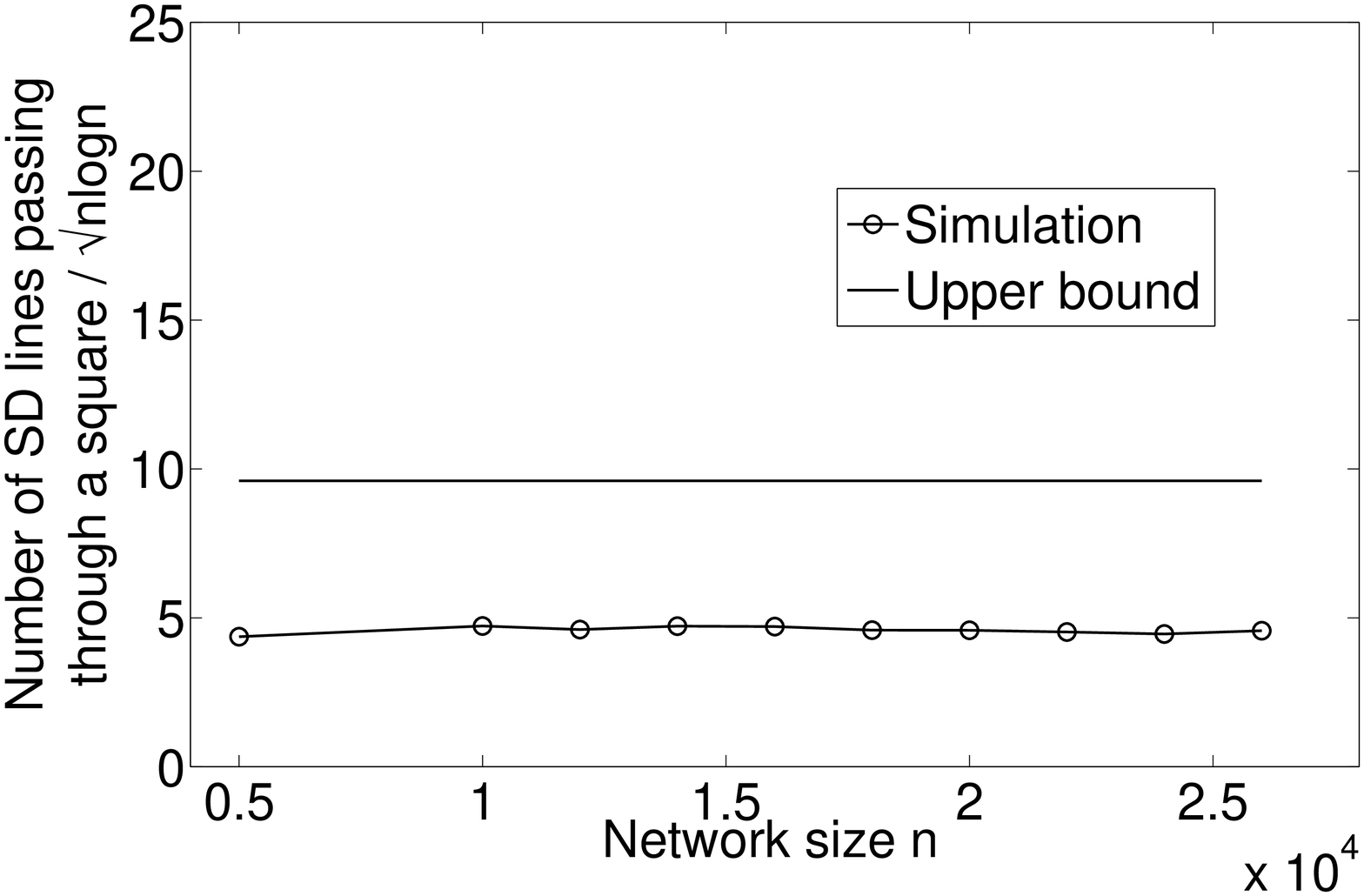}
\par\end{centering}

\protect\caption{\label{fig:number of SD lines}The number of SD lines passing through
a square versus the upper bound in Lemma \ref{lem:number of SD lines intersecting any square}.
Vertical axis shows the ratio of the number of SD lines passing through
a square to $\sqrt{n}\log n$.}
\end{figure}

Lemma \ref{lem:number of SD lines intersecting any square} shows
that the random number of SD lines passing through an arbitrarily
chosen square, including the SD lines originating from and ending
at the square, is upper bounded. 
\begin{lem}
\label{lem:number of SD lines intersecting any square}For an arbitrary
square in $B_{n}$, the random number of $SD$ lines passing through
it, denoted by $Y$, satisfies that
\begin{equation}
\lim_{n\rightarrow\infty}\Pr\left(Y\leq\omega_{2}\sqrt{n}\log n\right)=1\label{eq:bound on the number of SD line through a square}
\end{equation}
where $\omega_{2}=3.2\left(1+\epsilon\right)\left(1+\delta_{1}\right)c_{1}$,
$\epsilon$ and $\delta_{1}$ are arbitrarily small positive constants.\end{lem}
\begin{IEEEproof}
See Appendix I.
\end{IEEEproof}
As a way of establishing the tightness of the bound in Lemma \ref{lem:number of SD lines intersecting any square},
Fig. \ref{fig:number of SD lines} shows simulation results of the
number of SD lines passing a square in comparison with the upper bound
in Lemma \ref{lem:number of SD lines intersecting any square}. 

Using Corollary \ref{cor:number of vertical and horizontal paths}
and Lemma \ref{lem:number of SD lines intersecting any square}, the
following result can be readily established:
\begin{lem}
\label{lem:result on traffic load}Each relay node needs to carry
the traffic of at most $\frac{\omega_{2}\sqrt{n}}{0.5474}$ source-destination
pairs a.a.s.
\end{lem}
Note that a node not on an open path does not need to carry the traffic
of other source-destination pairs.

\section{A Solution to the Hidden-node Problem\label{sec:Hidden-node-free-design}}

\begin{figure}
\begin{centering}
\includegraphics[width=0.4\columnwidth]{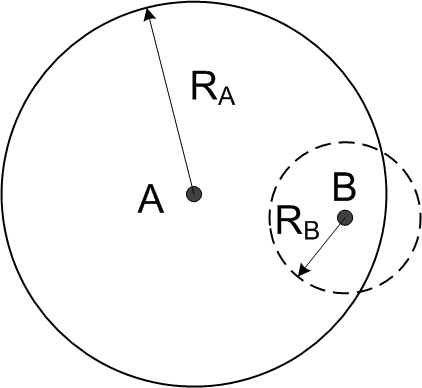}
\par\end{centering}

\protect\caption{An illustration of the hidden node problem when nodes use different
transmission power. Assume that the same carrier-sensing threshold
is used by node A and B. The transmission of A using a lager transmission
power (node B using a smaller transmission power, respectively) can
be detected by nodes located within a distance $R_{A}$ ($R_{B}$,
respectively), and $R_{A}>R_{B}$. Consequently B can detect A's transmission
but node A cannot detect node B's. Therefore even when node B is transmitting,
node A still can start its own transmission, thereby resulting in
a collision and causing the hidden node problem. \label{fig:Hidden node problem}}
\end{figure}

Our routing algorithm described in the last section needs to use two
different transmission ranges of lengths $\Theta\left(1\right)$ and
$\Theta\left(\log n\right)$ respectively. The use of two different
transmission ranges in CSMA networks will exacerbate the so-called
hidden node problem. See Fig. \ref{fig:Hidden node problem} for an
illustration. In \cite{Chau11Capacity}, the problem was addressed
by letting nodes operate on two frequency bands, namely, short-range
transmissions operate on one frequency band while long-range transmissions
operate on the other. Their solution may result in lower spectrum
usage because long-range transmission is used less frequently and
also poses additional hardware requirements on nodes. Therefore, we
present a solution by jointly tuning the transmission power and the
carrier-sensing threshold.

\subsection{A formal definition of the hidden node problem}

Under the SINR model, a set of concurrent transmissions (or links)
are said to form an \emph{independent set} if the SINRs are \emph{all}
above the SINR threshold $\beta$. Let $\mathcal{F}$ be the set of
all independent sets. Because of the random and distributed nature
of the carrier-sensing operations by individual nodes, the set of
simultaneous transmissions observing the carrier-sensing constraint,
denoted by $\mathcal{S}^{CS}$, may or may not belong to $\mathcal{F}$,
i.e., some transmissions observing the carrier-sensing constraints
may still cause the SINRs at some receivers to be above $\beta$.
Let $\mathcal{F}^{CS}$ be the set of all $\mathcal{S}^{CS}$s. Let
$\Psi$ be the set of concurrent transmissions in a CSMA network.
More formally, a hidden node problem is said to occur if $\Psi\in\mathcal{F}^{CS}$
but $\Psi\notin\mathcal{F}$. A\emph{ }CSMA network is said to be
\emph{hidden node free} if its carrier-sensing operations and transmission
powers are carefully designed such that all $\Psi\in\mathcal{F}^{CS}$
also meets the condition that $\Psi\in\mathcal{F}$.

For a CSMA network in which uniform transmission power is in use,
by setting the carrier-sensing range to be a constant multiple of
the transmission range, the hidden node problem can be effectively
eliminated \cite{Fu12Effective,Chau11Capacity}. For our routing algorithm
using two transmission ranges of lengths $\Theta\left(1\right)$ and
$\Theta\left(\log n\right)$ , if the carrier-sensing range is set
to be $\Theta\left(\log n\right)$, although the hidden node problem
can be eliminated, the number of concurrent transmissions (hence the
spatial frequency reuse) will be reduced compared with a carrier-sensing
range of $\Theta\left(1\right)$, which in turn causes a reduced capacity.
Therefore we manage to have transmissions with different ranges to
coexist concurrently instead. In this way, the capacity will be maximized
while eliminating the hidden node problem. 

More specifically, let $P_{i}$ be the transmission power used for
the $i^{th}$ transmission where the same transmitter may use different
power when transmitting to different receiver. The transmitter also
uses different carrier-sensing threshold when different transmission
power is used. Denote by $T_{i}$ the carrier-sensing threshold used
for $P_{i}$. Furthermore, let the transmission power of a transmitter
be such that the power received at its intended receiver is at least
$\bar{P}$ ($\bar{P}$ is a constant not depending on $n$ and the
value of $\bar{P}$ will be specified shortly later in this section).
The following lemma specifies the relation between $P_{i}$ and $T_{i}$
required for two transmitters to be able to sense each other's transmission.
\begin{lem}
\label{lem:mutually sense range}Let the values of $P_{i}$ and $T_{i}$
be chosen such that the following condition is met 
\begin{equation}
P_{i}=\bar{P}T_{i}^{-1}.\label{eq:relation between transmit power and carrier-sensing threshold}
\end{equation}
For two arbitrary transmitters located at $\boldsymbol{x}_{i}$ and
$\boldsymbol{x}_{j}$ respectively, they can sense each other's transmission
iff 
\begin{equation}
\left\Vert \boldsymbol{x}_{i}-\boldsymbol{x}_{j}\right\Vert <\left(\frac{\bar{P}}{T_{i}T_{j}}\right)^{\frac{1}{\alpha}}=\left(\frac{P_{i}P_{j}}{\bar{P}}\right)^{\frac{1}{\alpha}}\label{eq:mutual sensing distance among two nodes}
\end{equation}
\end{lem}
\begin{IEEEproof}
When node $i$ located at $\boldsymbol{x}_{i}$ transmits using power
$P_{i}$, the power received at node $j$ at location $\boldsymbol{x}_{j}$
is given by $P_{i}\left\Vert \boldsymbol{x}_{i}-\boldsymbol{x}_{j}\right\Vert ^{-\alpha}$.
Let $T_{j}$ be the carrier-sensing threshold of node $j$. The transmission
of node $i$ can be detected iff $P_{i}\left\Vert \boldsymbol{x}_{i}-\boldsymbol{x}_{j}\right\Vert ^{-\alpha}>T_{j}$.
Using (\ref{eq:relation between transmit power and carrier-sensing threshold}),
node $j$ can detect node $i$'s transmission iff $\bar{P}T_{i}^{-1}\left\Vert \boldsymbol{x}_{i}-\boldsymbol{x}_{j}\right\Vert ^{-\alpha}>T_{j}$
or equivalently $\left\Vert \boldsymbol{x}_{i}-\boldsymbol{x}_{j}\right\Vert <\left(\frac{\bar{P}}{T_{i}T_{j}}\right)^{\frac{1}{\alpha}}=\left(\frac{P_{i}P_{j}}{\bar{P}}\right)^{\frac{1}{\alpha}}$.
Using a similar argument, node $i$ can detect node $j$'s transmission
iff (\ref{eq:mutual sensing distance among two nodes}) is met.
\end{IEEEproof}
Lemma \ref{lem:mutually sense range} shows that by carefully choosing
the carrier-sensing threshold according to the transmission power
for each transmitter, a major cause of the hidden node problem: a
node $A$ senses another node $B$'s transmission but node $B$ cannot
sense node $A$'s transmission can be eliminated. In the next several
paragraphs, we shall demonstrate how to choose $\bar{P}$, which determines
the minimum power received at a receiver, such that the SINR requirement
can also be met.

In the first and third stages of our routing algorithm, the maximum
distance between a transmitter and a receiver is $\sqrt{2}c_{1}\log n$
while the the maximum distance between a transmitter and a receiver
in the second stage is $\sqrt{5}c$. Accordingly, for the first and
third stages, we let the transmission power be
\begin{equation}
P^{h}=\bar{P}\left(\sqrt{2}c_{1}\log n\right)^{\alpha}\label{eq:high transmit power}
\end{equation}
while for the second stage, the transmission power is set at 
\begin{equation}
P^{l}=\bar{P}\left(\sqrt{5}c\right)^{\alpha}\label{eq:low transmit power}
\end{equation}
It is trivial to show that the received signal power of all transmissions
is at least $\bar{P}$. Furthermore, the following theorem established
in our previous work \cite[Theorem 1]{Yang12Connectivity} helps to
obtain an upper bound on the interference experienced by any receiver
in the network.
\begin{thm}
\label{thm:interference homogeneous CSMA}Consider a CSMA network
with nodes distributed arbitrarily on a finite area in $\Re^{2}$
where all nodes transmit at the same power $P$ and use the same carrier-sensing
threshold $T$. Furthermore, the power received by a node at $\boldsymbol{x}_{j}$
from a transmitter at $\boldsymbol{x}_{i}$ is given by $P\left\Vert \boldsymbol{x}_{i}-\boldsymbol{x}_{j}\right\Vert ^{-\alpha}$.
Let $r_{0}$ be the distance between a receiver and its transmitter.
The maximum interference experienced by the receiver is smaller than
or equal to $N_{1}\left(d,r_{0}\right)+N_{2}\left(d\right)$ where
\begin{align}
N_{1}\left(d,r_{0}\right)= & \frac{4\left(\frac{5\sqrt{3}}{4}d-r_{0}\right)^{1-\alpha}\left(\frac{\sqrt{3}}{4}\left(3\alpha-1\right)d-r_{0}\right)}{d^{2}\left(\alpha-1\right)\left(\alpha-2\right)}\nonumber \\
+ & \frac{3}{\left(d-r_{0}\right)^{\alpha}}+\frac{3}{\left(\sqrt{3}d-r_{0}\right)^{\alpha}}+\frac{3\left(\frac{3}{2}d-r_{0}\right)^{1-\alpha}}{\left(\alpha-1\right)d}\label{eq:N_1}
\end{align}

\begin{eqnarray}
N_{2}\left(d\right) & = & \frac{3}{d^{\alpha}}+\frac{3(\frac{3}{2})^{1-\alpha}}{\left(\alpha-1\right)d^{\alpha}}+\frac{3}{\left(\sqrt{3}d\right)^{\alpha}}\nonumber \\
 &  & +\frac{3\left(\frac{5}{4}\right)^{1-\alpha}\left(3\alpha-1\right)}{\left(\alpha-1\right)\left(\alpha-2\right)\left(\sqrt{3}d\right)^{\alpha}}\label{eq:N_2}
\end{eqnarray}
and $d=\left(\frac{P}{T}\right)^{\frac{1}{\alpha}}$.
\end{thm}
Noting that $N_{1}\left(d,r_{0}\right)$ is a monotonically increasing
function of $r_{0}$, it can be readily established using Theorem
\ref{thm:interference homogeneous CSMA} that in the CSMA network
analyzed in this paper in which two sets of transmission powers, carrier-sensing
threshold and the maximum transmission range are employed, the maximum
interference (for any value of $n$) is bounded by
\begin{align}
 & N_{1}\left(\left(\frac{P^{h}}{T^{h}}\right)^{\frac{1}{\alpha}},\sqrt{2}c_{1}\log n\right)+N_{2}\left(\left(\frac{P^{h}}{T^{h}}\right)^{\frac{1}{\alpha}}\right)\nonumber \\
+ & N_{1}\left(\left(\frac{P^{l}}{T^{l}}\right)^{\frac{1}{\alpha}},\sqrt{5}c\right)+N_{2}\left(\left(\frac{P^{l}}{T^{l}}\right)^{\frac{1}{\alpha}}\right)\label{eq:interference for finite networks}
\end{align}
where $T^{l}$ and $T^{h}$ are the carrier-sensing threshold chosen
for $P^{l}$ and $P^{h}$ respectively according to (\ref{eq:relation between transmit power and carrier-sensing threshold}). 
\begin{rem}
At the expense of more analytical efforts, a tighter bound on interference
can be established that the maximum interference in the CSMA network
considered in this paper is bounded by $N_{1}\left(\left(\frac{P^{l}}{T^{l}}\right)^{\frac{1}{\alpha}},\sqrt{5}c\right)+N_{2}\left(\left(\frac{P^{l}}{T^{l}}\right)^{\frac{1}{\alpha}}\right)$
for \emph{any value} of $n$. Because for a sufficiently large network,
which is the focus of this paper, the difference between this bound
and the upper bound in (\ref{eq:interference for finite networks})
is negligibly small, we choose to omit the analysis due to space limitation.
\end{rem}
Noting that $\left(\frac{P^{h}}{T^{h}}\right)^{\frac{1}{\alpha}}=\bar{P}^{\frac{1}{\alpha}}\left(\sqrt{2}c_{1}\log n\right)^{2}$,
$\left(\frac{P^{l}}{T^{l}}\right)^{\frac{1}{\alpha}}=\bar{P}^{\frac{1}{\alpha}}\left(\sqrt{5}c\right)^{2}$
and $d=\left(\frac{P}{T}\right)^{\frac{1}{\alpha}}$, it is easy to
conclude using (\ref{eq:N_1}) and (\ref{eq:N_2}) that when $\alpha>1$,
the contribution of the first two terms $N_{1}\left(\left(\frac{P^{h}}{T^{h}}\right)^{\frac{1}{\alpha}},\sqrt{2}c_{1}\log n\right)+N_{2}\left(\left(\frac{P^{h}}{T^{h}}\right)^{\frac{1}{\alpha}}\right)$,
attributable to transmissions using a larger transmission power, become
vanishingly small compared with the last two terms as $n\rightarrow\infty$.
The following theorem provides guidance on how to choose $\bar{P}$
to meet the SINR requirements for all concurrent transmissions in
a large CSMA network.
\begin{thm}
\label{thm:hidden node free}For an arbitrarily high SINR requirement
$\beta$, there exists a value of $\bar{P}$ for sufficiently large
$n$ such that the SINR of all transmissions in a CSMA network, in
which each transmitter sets its transmission power and carrier sensing
threshold according to the relationship in Lemma \ref{lem:mutually sense range},
is greater than or equal to $\beta$. Furthermore, the value of $\bar{P}$
is given implicitly by the following equation
\begin{equation}
\frac{\bar{P}}{N_{1}\left(\left(\frac{P^{l}}{T^{l}}\right)^{\frac{1}{\alpha}},\sqrt{5}c\right)+N_{2}\left(\left(\frac{P^{l}}{T^{l}}\right)^{\frac{1}{\alpha}}\right)}=\beta\label{eq:power for hidden node free network}
\end{equation}
\end{thm}
\begin{IEEEproof}
Noting that the minimum received power is $\bar{P}$, the theorem
becomes an easy consequence of the interference upper bound established
earlier in the section.
\end{IEEEproof}
As a brief summary of the results of this section, Theorem \ref{thm:hidden node free}
gives guidance on how to choose $\bar{P}$ to meet the SINR requirement.
More specifically, noting that $\left(\frac{P^{l}}{T^{l}}\right)^{\frac{1}{\alpha}}=\bar{P}^{\frac{1}{\alpha}}\left(\sqrt{5}c\right)^{2}$
and using equations (\ref{eq:N_1}) and (\ref{eq:N_2}), equation
(\ref{eq:power for hidden node free network}) becomes an implicit
equation of $\bar{P}$. Solving the equation, the value of $\bar{P}$
that meets the SINR requirement can be obtained, which is independent
of $n$. Given the value of $\bar{P}$, the other parameters in the
CSMA network, i.e. $P^{h}$, $P^{l}$ and the carrier sensing thresholds,
can all be determined using equations (\ref{eq:relation between transmit power and carrier-sensing threshold}),
(\ref{eq:high transmit power}) and (\ref{eq:low transmit power})
respectively. It can be readily established using the analysis presented
in this section that the CSMA network whose transmission power and
carrier sensing threshold are chosen following the above steps are
immune from the hidden node problem.

\section{Back Off Timer Setting and Capacity Analysis\label{sec:Capacity-Analysis}}

In the last section, we demonstrated how to choose the transmission
power and the carrier sensing threshold to solve the hidden node problem.
In the CSMA network in which nodes may use two different transmission
powers, a potential problem that may arise is that nodes using the
larger transmission power may potentially contend with more nodes
for transmission opportunities. Therefore nodes using the larger transmission
power may not get a fair transmission opportunity compared with nodes
using the smaller transmission power. This may potentially causes
nodes using the larger transmission power to become a bottleneck in
throughput which reduces the overall network capacity. In this section,
we demonstrate how to choose another controllable parameter in CSMA
protocols, i.e., back off timer, to conquer the difficulty.

Same as that in references \cite{Chau11Capacity} and \cite{Jiang10A},
we consider a CSMA protocol in which the initial back off timer is
a random variable following an exponential distribution. Nodes using
different transmission power may however choose different mean value
to use in the exponential distribution governing their respective
random initial back off timer. The following theorem provides the
basis for choosing these mean values.

\begin{thm}
\label{thm:medium access probability}Let $\delta_{2}$ and $\delta_{3}$
be two small positive constants. If transmissions using a low transmit
power $P^{l}$ set their initial back off time to be exponentially
distributed with mean $\lambda_{l}=1$ and transmissions using a high
transmission power $P^{h}$ set their initial back off time to be
exponentially distributed with mean $\lambda_{h}=\frac{1}{\log^{2}n}$,
then 

(i) a.a.s. each low power transmission can be active with a constant
probability greater than or equal to 
\begin{equation}
\omega_{3}=\frac{1}{\pi\left(5\bar{P}^{\frac{1}{\alpha}}c\right)^{2}+\left(1+\delta_{2}\right)10\pi c^{2}c_{1}^{2}\bar{P}^{\frac{2}{\alpha}}+1}\label{eq:omega_3}
\end{equation}

(ii) a.a.s. each high power transmission can be active with a probability
greater than or equal to 
\[
\omega_{4}=\frac{1}{\pi\left(\sqrt{10}c_{1}\bar{P}^{\frac{1}{\alpha}}\right)^{2}\log^{4}n+\left(1+\delta_{3}\right)4\pi c_{1}^{4}\bar{P}^{\frac{2}{\alpha}}\log^{4}n+1}
\]
\end{thm}
\begin{IEEEproof}
See Appendix II.
\end{IEEEproof}
Fig. \ref{fig:medium access probability} shows the transmission opportunity
(or the medium access probability) of a node using $P^{l}$ versus
the lower bound in Theorem \ref{thm:medium access probability} for
different values of $n$.

\begin{figure}
\begin{centering}
\includegraphics[width=1\columnwidth]{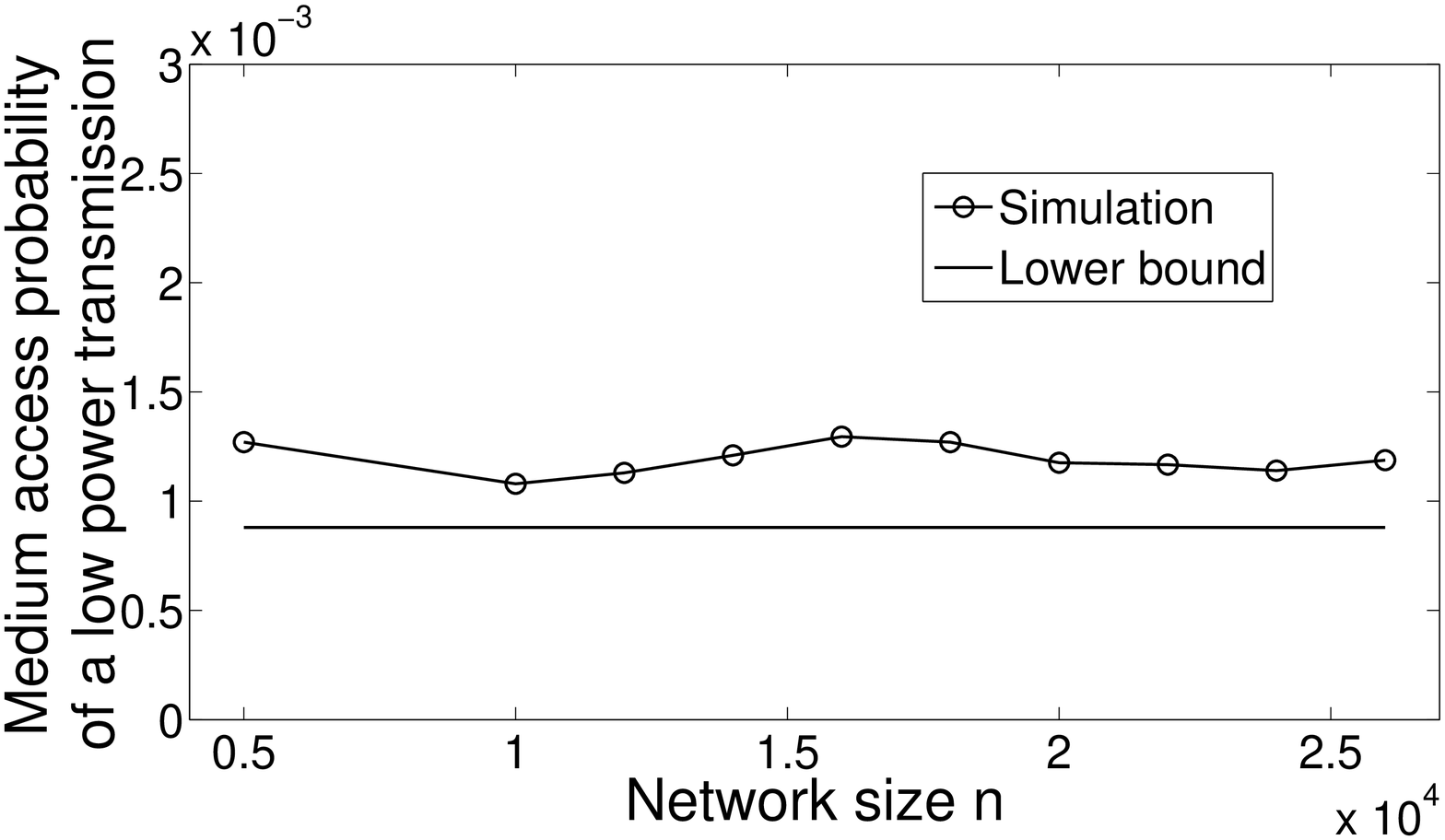}
\par\end{centering}

\protect\caption{\label{fig:medium access probability}A comparison between the simulation
result on the medium access probability of a node using the low power
transmission with the lower bound in Theorem \ref{thm:medium access probability}
where $\beta=10$ and $\alpha=4$. }
\end{figure}

On the basis of the results established in this section and in the
earlier sections, we present the following theorem which forms the
major result of this paper. 
\begin{thm}
\textbf{\label{thm:Throughput-lower-bound}}The achievable per-node
throughput in the CSMA network is greater than or equal to 
\begin{equation}
\frac{0.5474\omega_{3}}{\omega_{2}\sqrt{n}}W;\label{eq:pre-constant}
\end{equation}

and is smaller than or equal to 
\[
\frac{1}{0.52c\left(5\pi c^{2}c_{1}^{2}\bar{P}^{\frac{2}{\alpha}}+1\right)\sqrt{n}}\times W
\]
a.a.s. as $n\rightarrow\infty$, where $\omega_{2}$ is given in Lemma
\ref{lem:number of SD lines intersecting any square} and $\omega_{3}$
is given by (\ref{eq:omega_3}).\end{thm}
\begin{IEEEproof}
We first show that the achievable per-node throughput is lower bounded
by $\frac{0.5474\omega_{3}}{\omega_{2}\sqrt{n}}W$. Let $\lambda_{1}\left(n\right)$
($\lambda_{2}\left(n\right)$, respectively) be the per-node throughput
that can be achieved in the first and the third (the second, respectively)
stages of our routing algorithm. Obviously the final per-node throughput
$\lambda\left(n\right)$ satisfies $\lambda\left(n\right)=\min\left\{ \lambda_{1}\left(n\right),\lambda_{2}\left(n\right)\right\} $.
In the following, we analyze $\lambda_{1}\left(n\right)$ and $\lambda_{2}\left(n\right)$
separately.

As an easy consequence of Lemma \ref{lem:result on traffic load},
a.a.s. each relay node carries the traffic of at most $\frac{\omega_{2}\sqrt{n}}{0.5474}$
source-destination pairs. According to the first statement of Theorem
\ref{thm:medium access probability}, a.a.s. each relay node on an
open path can access the channel with a probability of at least $\omega_{3}$,
which is a constant independent of $n$. The conclusion then readily
follows that $\lim_{n\rightarrow\infty}\Pr\left(\lambda_{2}\left(n\right)\geq\frac{0.5474\omega_{3}}{\omega_{2}\sqrt{n}}W\right)=1$.

For the second stage of the routing, note that a source or a destination
node \emph{not} on an open path does not need to carry traffic for
other source-destination pairs. Using the second statement of Theorem
\ref{thm:medium access probability}, conclusion follows that $\lambda_{1}\left(n\right)=\Omega\left(\frac{1}{\log^{4}n}\right)$.

Combining the above two results on $\lambda_{1}\left(n\right)$ and
$\lambda_{2}\left(n\right)$ and noting that the capacity bottleneck
lies in the second stage, the first statement in this theorem is proved.

We now further  show that the achievable per-node throughput is upper
bounded by $\frac{W}{0.52c\left(5\pi c^{2}c_{1}^{2}\bar{P}^{\frac{2}{\alpha}}+1\right)\sqrt{n}}$.
The upper bound is to be established using a result proved in our
previous work \cite[Corollary 6]{Mao13Towards}, which shows that
the per-node throughput is equal to the product of the average number
of simultaneous transmissions and the link capacity divided by the
product of the average number of transmissions required to deliver
a packet to its destination and the number of source-destination pairs.
We first analyze the average number of transmissions required for
a packet to reach its destination. The average distance between a
randomly chosen source-destination pair is $0.52\sqrt{n}$ \cite{Philip07The}.
A packet moves by one cell in each hop on an open path where the contribution
of the last mile transmission between a source (a destination) and
an open-path node is vanishingly small compared with $0.52\sqrt{n}$.
Thus a.a.s. the average number of hops traversed by a packet is at
least $\frac{0.52\sqrt{n}}{c}$. Next we analyze the average number
of simultaneous transmissions. Since there is at most one node in
a cell acting as an open path node, there are at most $\frac{n}{c^{2}}$
open path nodes in the network. Following the same procedure in obtaining
(\ref{eq:number of conflicting nodes}), (\ref{eq:MAP by definition})
and (\ref{eq:transmission opportunities of low power links}), we
have that $\Pr\left[\eta_{i}^{l}\right]\leq\frac{1}{5\pi c^{2}c_{1}^{2}\bar{P}^{\frac{2}{\alpha}}+1}$.
Therefore, the average number of simultaneous transmissions is at
most $\frac{n}{c^{2}}\times\frac{1}{5\pi c^{2}c_{1}^{2}\bar{P}^{\frac{2}{\alpha}}+1}$
(Note that when a non-open-path node transmits using $P^{h}$, the
number of simultaneous transmissions will only reduce). As a ready
consequence of the above analysis and \cite[Corollary 6]{Mao13Towards},
an upper bound on the per-node throughput results.
\end{IEEEproof}
The lower bound on the per-node throughput provided in Theorem \ref{thm:Throughput-lower-bound}
is order optimal in the sense that the throughput is of the same order
as the known result on the optimum per-node throughput \cite{Franceschetti07Closing}
of networks under the same settings. Furthermore, Theorem \ref{thm:Throughput-lower-bound}
gives the pre-constant preceding the order of the per-node throughput:
$\frac{0.5474\omega_{3}}{\omega_{2}}$. A detail examination of the
pre-constant reveals that the pre-constant can be separated into the
product of two terms: $\frac{0.5474}{\omega_{2}}$ and $\omega_{3}$.
The first term $\frac{0.5474}{\omega_{2}}$ is entirely determined
by the routing algorithm, more specifically determined by how the
routing algorithm distribute traffic load among relay nodes and among
source-destination pairs. The second term $\omega_{3}$ is entirely
determined by the scheduling algorithm and some physical layer details,
i.e., the SINR requirement, interference and propagation model. The
above observation appears to suggest that impact of the routing algorithm
and the scheduling algorithm can be decoupled and studied separately,
and the two algorithms that determine the overall network capacity
can be optimized separately.

\begin{figure}
\begin{centering}
\includegraphics[width=1\columnwidth]{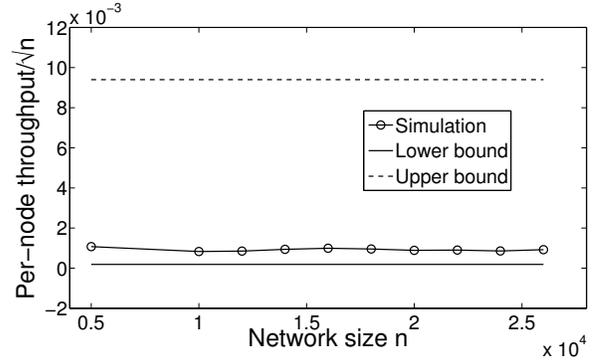}
\par\end{centering}

\protect\caption{\label{fig:throughput}A simulation of per-node throughput with $\alpha=4$
and $\beta=10$. For comparison, the upper and the lower bound obtained
in the paper is also shown.}
\end{figure}

\begin{figure}
\begin{centering}
\includegraphics[width=1\columnwidth]{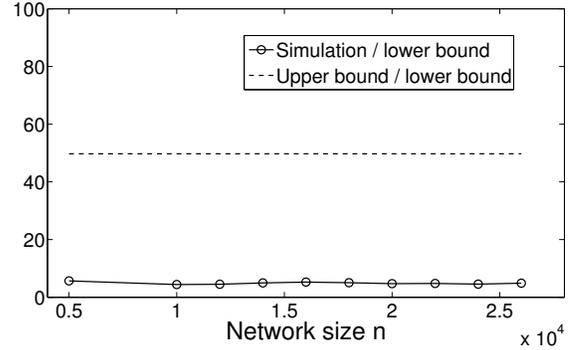}
\par\end{centering}

\protect\caption{\label{fig:Tightness of two bounds}A simulation of per-node throughput
with $\alpha=4$ and $\beta=10$. For comparison, the upper and the
lower bound obtained in the paper are also shown. To facilitate comparison,
both the per-node throughput obtained from simulations and the per-node
throughput upper bound are normalized by the per-node throughput lower
bound.}
\end{figure}

Fig. \ref{fig:throughput} shows a comparison of the per-node throughput
obtained from simulations, the upper lower bounds obtained in Theorem
\ref{thm:Throughput-lower-bound} for different values of $n$. To
facilitate comparison, Fig. \ref{fig:Tightness of two bounds} further
shows the ratio of the per-node throughput obtained from simulations
to the throughput lower bound and the ratio of the throughput upper
bound to the throughput lower bound. As shown in the figures, the
lower bound is fairly tight and the upper bound is also within a factor
of $10$ of the simulation result. The simulation results demonstrate
that the pre-constant obtained in our study provides a pretty accurate
characterization of the per-node throughput.

\section{Conclusion\label{sec:Conclusion}}

In this paper, we studied the transport capacity of large wireless
multi-hop CSMA networks. We showed that by carefully choosing the
controllable parameters in the CSMA protocol and designing the routing
algorithm, a network running distributed CSMA scheduling algorithm
and each node making routing decisions based on local information
only can also achieve an order-optimal throughput of $\Theta\left(\frac{1}{\sqrt{n}}\right)$,
which is the same as that of large networks employing centralized
routing and scheduling algorithms. Furthermore, we not only gave the
order of the throughput but also derived the pre-constant preceding
the order by giving an upper and a lower bound of the transport capacity.
The tightness of the bounds was validated using simulations. Theoretical
analysis was presented on tuning the carrier-sensing threshold and
the transmission power to avoid hidden node problems and on tuning
the back off timer distribution to ensure each node gain a fair access
to the channel in CSMA networks using non-uniform transmission powers.
The principle developed through the analysis was expected to be also
helpful to set the corresponding parameters of CSMA networks in a
more realistic setting.

\section*{Appendix I Proof of Lemma \ref{lem:number of SD lines intersecting any square}}

In the proof of Lemma \ref{lem:number of SD lines intersecting any square},
we will make use of a result established in the stochastic ordering
theory \cite{Klenke10Stochastic}. For two real valued random variables
$X_{1}$ and $X_{2}$, we say $X_{1}\leq_{st}X_{2}$ iff for all $x\in\left(-\infty,\infty\right)$,
$\Pr\left(X_{1}>x\right)\leq\Pr\left(X_{2}>x\right)$. 
\begin{thm}
\label{thm:Stochastic Ordering}\cite[Theorem 1(a)]{Klenke10Stochastic}Suppose
$X_{i}$ follows a Binomial distribution with parameters $n_{i}\in\mathbb{N}$
and $p_{i}\in\left(0,1\right)$, denote the distribution of $X_{i}$
by $B\left(n_{i},p_{i}\right)$, $i=1,2$, i.e., $X_{i}\sim B\left(n_{i},p_{i}\right)$.
We have $X_{1}\leq_{st}X_{2}$ iff $\left(1-p_{1}\right)^{n_{1}}\geq\left(1-p_{2}\right)^{n_{2}}$
and $n_{1}\leq n_{2}$. 
\end{thm}
As an easy consequence of the above theorem, for three independent
Binomial random variables $X_{1}\sim B\left(n_{1},p_{1}\right)$,
$X_{2}\sim B\left(n_{1},p_{2}\right)$ and $X_{3}\sim B\left(n_{2},p_{2}\right)$
with $n_{1}\leq n_{2}$ and $p_{1}\leq p_{2}$, it can be concluded
that $X_{1}\leq_{st}X_{2}\leq_{st}X_{3}$. 

\begin{figure}
\begin{centering}
\includegraphics[width=0.7\columnwidth]{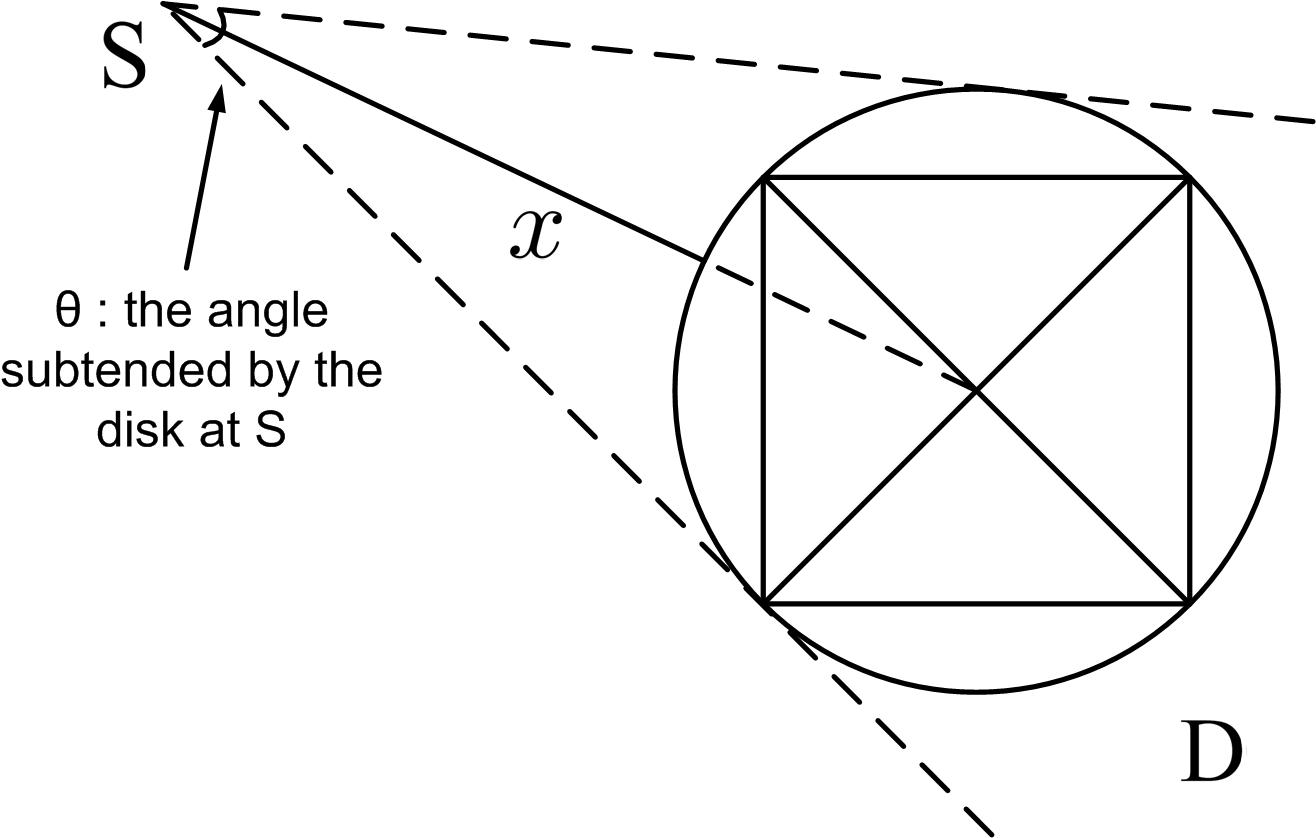}
\par\end{centering}

\protect\caption{\label{fig:SD_1}An illustration of a SD line intersecting the circumscribed
circle}
\end{figure}

Now we are ready to prove Lemma \ref{lem:number of SD lines intersecting any square}.
Let $Y_{i}^{j}$ be the indicator random variable for the event that
the $SD_{i}$ passes through the $j^{th}$ square:
\[
Y_{i}^{j}=\begin{cases}
1 & \textrm{if \ensuremath{SD_{i}}passes through the \ensuremath{j}th square}\\
0 & \textrm{otherwise.}
\end{cases}
\]
We shall derive an upper bound on $\Pr\left[Y_{i}^{j}=1\right]$ for
any $j\in\left[1,\,\frac{n}{\log^{2}n}\right]$. Circumscribe the
$j^{th}$ square with a small circle of radius $\frac{\sqrt{2}}{2}c_{1}\log n$,
as shown in Fig. \ref{fig:SD_1}. For a source $S$ located outside
the square and at a distance $x$ from the center of the square, the
angle $\theta\left(x\right)$ subtended by the circle at S is $\theta\left(x\right)=2\arcsin\frac{\frac{\sqrt{2}}{2}c_{1}\log n}{x}$.
Using the fact that $\arcsin x\leq1.6x$ when $0\leq x\leq1$, we
have 
\begin{equation}
\theta\left(x\right)=1.6\arcsin\frac{\frac{\sqrt{2}}{2}c_{1}\log n}{x}\leq3.2\frac{\frac{\sqrt{2}}{2}c_{1}\log n}{x}\label{eq:upper bound on theta}
\end{equation}

Noting that $B_{n}$ is of size $\sqrt{n}\times\sqrt{n}$, the area
of the sector formed by the two dashed tangents Fig. \ref{fig:SD_1}
and the boarder of $B_{n}$ is at most $\frac{\theta\left(x\right)}{2\pi}n$.
If the destination of $S$, denoted by $D$, does not lie in this
sector, then the associated $SD$ line does not pass through the circle.
Therefore, the probability that the $SD$ line intersecting the circle
is at most $\frac{\theta\left(x\right)}{2\pi}$. Considering that
the circle is located in a $\sqrt{n}\times\sqrt{n}$ box $B_{n}$,
the probability density that $S$ is at a distance $x$ from the circle
can be shown to be upper bounded by $\frac{2\pi x}{n}$. It follows
from the above analysis and (\ref{eq:upper bound on theta}) that
\begin{align}
 & \Pr\left[Y_{i}^{j}=1\right]\nonumber \\
\leq & \int_{0}^{\sqrt{2n}}\frac{3.2\times\frac{\sqrt{2}}{2}c_{1}\log n}{2\pi x}\times\frac{2\pi x}{n}\mathrm{d}x=\frac{3.2c_{1}\log n}{\sqrt{n}}\label{eq:Pr=00005BS-D intersects a square=00005D}
\end{align}

Recall that $\Gamma$ represents the set of indices of all nodes in
the network. For a fixed square $j$, the total number of $SD$ lines
passing through it is given by $Y^{j}=\sum_{i=1}^{\left|\Gamma\right|}Y_{i}^{j}$,
which is the sum of i.i.d. Bernoulli random variables since the locations
of nodes are independent and $Y_{i}^{j}$ depends only on the locations
of source and destination nodes of the $i^{th}$ source-destination
pair. Therefore $Y^{j}$ follows the Binomial distribution, i.e.,
$Y^{j}\sim B\left(\left|\Gamma\right|,\,\Pr\left[Y_{i}^{j}=1\right]\right)$.
As an easy consequence of the Poisson distribution of nodes, a.a.s.
the total number of nodes $\left|\Gamma\right|\leq\left(1+\epsilon\right)n$,
where $\epsilon$ is an arbitrarily small positive constant. Define
another Binomial random variable $\tilde{Y}^{j}\sim B\left(\left(1+\epsilon\right)n,\frac{4c_{1}\log n}{\sqrt{n}}\right)$
. It follows from Theorem \ref{thm:Stochastic Ordering} that
\[
Y^{j}\leq_{\textrm{st}}\tilde{Y}^{j}
\]
It can be further shown that for any $0<\delta_{1}<1$,
\begin{eqnarray}
 &  & \Pr\left[Y^{j}>\left(1+\delta_{1}\right)\left(1+\epsilon\right)n\frac{3.2c_{1}\log n}{\sqrt{n}}\right]\nonumber \\
 & \leq & \Pr\left[\tilde{Y}^{j}>\left(1+\delta_{1}\right)\left(1+\epsilon\right)n\frac{3.2c_{1}\log n}{\sqrt{n}}\right]\nonumber \\
 & = & \Pr\left[\tilde{Y}^{j}>\left(1+\delta_{1}\right)\textrm{E}\left[\tilde{Y}^{j}\right]\right]\nonumber \\
 & \leq & \exp\left(-\frac{\delta_{1}^{2}}{3}\textrm{E}\left[\tilde{Y}^{j}\right]\right)\label{eq:Chernoff-1}\\
 & = & \exp\left(-\frac{3.2\left(1+\epsilon\right)\delta_{1}^{2}c_{1}\sqrt{n}\log n}{3}\right)
\end{eqnarray}
where (\ref{eq:Chernoff-1}) results from the Chernoff bound. Using
the union bound and the above result, we have
\begin{eqnarray}
 &  & \Pr\left[\bigcup_{j=1}^{\frac{n}{c_{1}^{2}\log^{2}n}}Y^{j}>3.2\left(1+\epsilon\right)\left(1+\delta_{1}\right)c_{1}\sqrt{n}\log n\right]\nonumber \\
 & \leq & \frac{n}{c_{1}^{2}\log^{2}n}\exp\left(-\frac{3.2\left(1+\epsilon\right)\delta_{1}^{2}c_{1}\sqrt{n}\log n}{3}\right)\label{eq:union bound-1}
\end{eqnarray}
Noting that $\frac{n}{c_{1}^{2}\log^{2}n}\exp\left(-\frac{3.2\left(1+\epsilon\right)\delta_{1}^{2}c_{1}\sqrt{n}\log n}{3}\right)\rightarrow0$
as $n\rightarrow\infty$, therefore a.a.s. $Y^{j}\leq3.2\left(1+\epsilon\right)\left(1+\delta_{1}\right)c_{1}\sqrt{n}\log n$
for any $j\in\left[1,\,\frac{n}{\log^{2}n}\right]$ which completes
the proof of Lemma \ref{lem:number of SD lines intersecting any square}.

\section*{Appendix II Proof of Theorem \ref{thm:medium access probability}}

Consider a node $i$ on an open path located at $\boldsymbol{x}_{i}$
transmitting with power $P^{l}=\bar{P}\left(\sqrt{5}c\right)^{\alpha}$.
Since the highest transmission power used in the network is $P^{h}=\bar{P}\left(\sqrt{2}c_{1}\log n\right)^{\alpha}$,
by (\ref{eq:mutual sensing distance among two nodes}), the furtherest
transmitter that node $i$ can sense is within a distance of $\sqrt{10}cc_{1}\bar{P}^{\frac{1}{\alpha}}\log n$.
Denote by $\mathcal{D}\left(\boldsymbol{x},\, r\right)$ a disk centered
at $\boldsymbol{x}$ and with a radius of $r$. All nodes that are
possibly competing with node $i$ for transmission opportunities are
located within $\mathcal{D}\left(\boldsymbol{x}_{i},\,\sqrt{10}cc_{1}\bar{P}^{\frac{1}{\alpha}}\log n\right)$.
Denote by $\mathcal{A}\left(\boldsymbol{x},\, r_{1},\, r_{2}\right)$
an annulus area centered at $\boldsymbol{x}$ with an inner radius
$r_{1}$ and an outer radius $r_{2}$. A little reflection shows that
all nodes using the low transmission power $P^{l}$ \emph{and} competing
with node $i$ must be located in $\mathcal{D}\left(\boldsymbol{x}_{i},\,5\bar{P}^{\frac{1}{\alpha}}c^{2}\right)$,
and the nodes in $\mathcal{A}\left(\boldsymbol{x}_{i},\,5\bar{P}^{\frac{1}{\alpha}}c^{2},\,\sqrt{10}cc_{1}\bar{P}^{\frac{1}{\alpha}}\log n\right)$
that compete with node $\boldsymbol{x}_{i}$ must use the high transmit
power $P^{h}$. Note that in each open site that forms the open path,
only one node serves as the relay node. Hence, there are at most $\frac{\pi\left(5\bar{P}^{\frac{1}{\alpha}}c^{2}\right)^{2}}{c^{2}}=\pi\left(5\bar{P}^{\frac{1}{\alpha}}c\right)^{2}$
open path nodes in $\mathcal{D}\left(\boldsymbol{x}_{i},\,5\bar{P}^{\frac{1}{\alpha}}c^{2}\right)$
that use $P^{l}$. Let $\mathcal{N}\left(\boldsymbol{x},\, r\right)$
be the random number of nodes located in $\mathcal{D}\left(\boldsymbol{x},\, r\right)$.
Next we provide an asymptotic upper bound on the number of  nodes
in $\mathcal{D}\left(\boldsymbol{x}_{i},\,\sqrt{10}cc_{1}\bar{P}^{\frac{1}{\alpha}}\log n\right)$
for any node $i$ on an open path. Denoting by $\mathcal{H}$ the
set of indices of nodes on open paths, clearly $\left|\mathcal{H}\right|<\frac{n}{c^{2}}$.
By Chernoff bound and the union bound, we have for an arbitrarily
small positive constant $\delta_{2}$,
\begin{eqnarray}
 &  & \Pr\left[\bigcup_{i\in\mathcal{H}}\mathcal{N}\left(\boldsymbol{x}_{i},\sqrt{10}cc_{1}\bar{P}^{\frac{1}{\alpha}}\log n\right)\geq\right.\nonumber \\
 &  & \left.\left(1+\delta_{2}\right)10\pi c^{2}c_{1}^{2}\bar{P}^{\frac{2}{\alpha}}\log^{2}n\right]\nonumber \\
 & = & \Pr\left[\bigcup_{i\in\mathcal{H}}\mathcal{N}\left(\boldsymbol{x}_{i},\,\sqrt{10}cc_{1}\bar{P}^{\frac{1}{\alpha}}\log n\right)\geq\right.\nonumber \\
 &  & \left.\left(1+\delta_{2}\right)\textrm{E}\left[\mathcal{N}\left(\boldsymbol{x}_{i},\,\sqrt{10}cc_{1}\bar{P}^{\frac{1}{\alpha}}\log n\right)\right]\right]\nonumber \\
 & \leq & \frac{n}{c^{2}}e^{-\frac{\delta_{2}^{2}}{3}\textrm{E}\left[\mathcal{N}\left(\boldsymbol{x}_{i},\,\sqrt{10}cc_{1}\bar{P}^{\frac{1}{\alpha}}\log n\right)\right]}\label{eq:number of conflicting nodes}
\end{eqnarray}
where $E$ denotes the expectation operator. It can be readily shown
that $\frac{n}{c^{2}}\exp\left\{ -\frac{\delta_{2}^{2}}{3}\textrm{E}\left[\mathcal{N}\left(\boldsymbol{x}_{i},\,\sqrt{10}cc_{1}\bar{P}^{\frac{1}{\alpha}}\log n\right)\right]\right\} $
approaches $0$ as $n\rightarrow\infty$. Therefore a.a.s. the number
of nodes within a distance $\sqrt{10}cc_{1}\bar{P}^{\frac{1}{\alpha}}\log n$
of an open path node is bounded above by $\left(1+\delta_{2}\right)10\pi c^{2}c_{1}^{2}\bar{P}^{\frac{2}{\alpha}}\log^{2}n$.

Next we analyze the transmission opportunity of an open path node.
Denote by $t_{i}$ the back off timer of node $i$ at a particular
time instant when the channel is idle. Denote by $\mathcal{C}_{i}$
the set of indices of nodes that compete with node $i$ for transmission.
Following the CSMA protocol, node $i$ can become an active transmitter
in the competition if 
\[
t_{i}<\min_{j\in\mathcal{C}_{i}\backslash\left\{ i\right\} }t_{j}.
\]
Let $\eta_{i}^{l}$ be the event that a transmission of node $i$
using the low transmit power is active. Using the ``memoryless''
property of an exponential distribution that for a timer following
an exponential distribution, the amount of lapsed time does not alter
the distribution of the remaining value of the timer, it can be shown
that for any $i\in\mathcal{H}$
\begin{eqnarray}
 &  & \Pr\left[\eta_{i}^{l}\right]\nonumber \\
 & = & \Pr\left[t_{i}<\min_{j\in\mathcal{C}_{i}\backslash\left\{ i\right\} }t_{j}\right]\nonumber \\
 & \geq & \int_{0}^{\infty}\left(e^{-\lambda_{l}t}\right)^{\pi\left(5\bar{P}^{\frac{1}{\alpha}}c\right)^{2}}\left(e^{-\lambda_{h}t}\right)^{\left(1+\delta_{2}\right)10\pi c^{2}c_{1}^{2}\bar{P}^{\frac{2}{\alpha}}\log^{2}n}\nonumber \\
 &  & \times\lambda_{l}e^{-\lambda_{l}t}\mathrm{d}t\label{eq:MAP by definition}
\end{eqnarray}
where in the above equation the term $\left(e^{-\lambda_{l}t}\right)^{\pi\left(5\bar{P}^{\frac{1}{\alpha}}c\right)^{2}}$
represents the probability that at a randomly chosen time instant
when the channel is idle, all $\pi\left(5\bar{P}^{\frac{1}{\alpha}}c\right)^{2}$
open path nodes in $\mathcal{D}\left(\boldsymbol{x}_{i},\,5\bar{P}^{\frac{1}{\alpha}}c^{2}\right)$,
which are competing for transmission opportunities with node $i$,
have their respective back off timer larger than a particular value
$t$; the term $\left(e^{-\lambda_{h}t}\right)^{\left(1+\delta_{2}\right)10\pi c^{2}c_{1}^{2}\bar{P}^{\frac{2}{\alpha}}\log^{2}n}$
represents the probability that all nodes using $P^{h}$ in $\mathcal{D}\left(\boldsymbol{x}_{i},\,\sqrt{10}cc_{1}\bar{P}^{\frac{1}{\alpha}}\log n\right)$,
which are competing for transmission opportunities with node $i$,
have their respective back off timer larger than $t$; the term $\lambda_{l}e^{-\lambda_{l}t}$
is the pdf of the back off timer of node $i$. It can be further shown
from (\ref{eq:MAP by definition}) that for any $i\in\mathcal{H}$,
\begin{eqnarray}
 &  & \Pr\left[\eta_{i}^{l}\right]\nonumber \\
 & \geq & \lambda_{l}\int_{0}^{\infty}e^{-\left(\pi\left(5\bar{P}^{\frac{1}{\alpha}}c\right)^{2}\lambda_{l}+\lambda_{h}\left(1+\delta_{2}\right)10\pi c^{2}c_{1}^{2}\bar{P}^{\frac{2}{\alpha}}\log^{2}n+\lambda_{l}\right)t}\mathrm{d}t\nonumber \\
 & = & \frac{\lambda_{l}}{\pi\left(5\bar{P}^{\frac{1}{\alpha}}c\right)^{2}\lambda_{l}+\lambda_{h}\left(1+\delta_{2}\right)10\pi c^{2}c_{1}^{2}\bar{P}^{\frac{2}{\alpha}}\log^{2}n+\lambda_{l}}\nonumber \\
 & = & \frac{1}{\pi\left(5\bar{P}^{\frac{1}{\alpha}}c\right)^{2}+\frac{\lambda_{h}}{\lambda_{l}}\left(1+\delta_{2}\right)10\pi c^{2}c_{1}^{2}\bar{P}^{\frac{2}{\alpha}}\log^{2}n+1}\nonumber \\
 & = & \frac{1}{\pi\left(5\bar{P}^{\frac{1}{\alpha}}c\right)^{2}+\left(1+\delta_{2}\right)10\pi c^{2}c_{1}^{2}\bar{P}^{\frac{2}{\alpha}}+1}.\label{eq:transmission opportunities of low power links}
\end{eqnarray}

Now we continue to prove the second part of Theorem \ref{thm:medium access probability}.
Consider that a node $j$ transmits using the high power $P^{h}=\bar{P}\left(\sqrt{2}c_{1}\log n\right)^{\alpha}$.
By (\ref{eq:mutual sensing distance among two nodes}), all nodes
that are possibly competing with node $j$ are located within $\mathcal{D}\left(\boldsymbol{x}_{j},\,2c_{1}^{2}\bar{P}^{\frac{1}{\alpha}}\log^{2}n\right)$.
Furthermore, among the nodes competing with node $j$, those open
path nodes using the lower transmission power $P^{l}$ must be located
in $\mathcal{D}\left(\boldsymbol{x}_{j},\,\sqrt{10}cc_{1}\bar{P}^{\frac{1}{\alpha}}\log n\right)$,
and the number of these open path nodes is \emph{at most} $\frac{\pi\left(\sqrt{10}cc_{1}\bar{P}^{\frac{1}{\alpha}}\log n\right)^{2}}{c^{2}}=\pi\left(\sqrt{10}c_{1}\bar{P}^{\frac{1}{\alpha}}\log n\right)^{2}$.
Next we derive an upper bound on the number of nodes in $\mathcal{D}\left(\boldsymbol{x}_{j},\,2c_{1}^{2}\bar{P}^{\frac{1}{\alpha}}\log^{2}n\right)$
competing with node $j$ for any $j\in\mathcal{O}$ where $\mathcal{O}$
is the set of indices of nodes using the high power. It can be easily
shown that $\lim_{n\rightarrow\infty}\Pr\left(\left|\mathcal{O}\right|<2n\right)=1$.
Using the union bound and the Chernoff bound, we have for any small
positive constant $\delta_{3}$,
\begin{eqnarray*}
 &  & \Pr\left[\bigcup_{j\in\mathcal{O}}\mathcal{N}\left(\boldsymbol{x}_{j},2c_{1}^{2}\bar{P}^{\frac{1}{\alpha}}\log^{2}n\right)\geq\right.\\
 &  & \left.\left(1+\delta_{3}\right)4\pi c_{1}^{4}\bar{P}^{\frac{2}{\alpha}}\log^{4}n\right]\\
 & = & \Pr\left[\bigcup_{j\in\mathcal{O}}\mathcal{N}\left(\boldsymbol{x}_{j},2c_{1}^{2}\bar{P}^{\frac{1}{\alpha}}\log^{2}n\right)\geq\right.\\
 &  & \left.\left(1+\delta_{3}\right)\textrm{E}\left[\mathcal{N}\left(\boldsymbol{x}_{j},2c_{1}^{2}\bar{P}^{\frac{1}{\alpha}}\log^{2}n\right)\right]\right]\\
 & \leq & 2ne^{-\frac{\delta_{3}^{2}}{3}\textrm{E}\left[\mathcal{N}\left(\boldsymbol{x}_{j},\,2c_{1}^{2}\bar{P}^{\frac{1}{\alpha}}\log^{2}n\right)\right]}
\end{eqnarray*}
Obviously $2n\exp\left\{ -\frac{\delta_{3}^{2}}{3}\textrm{E}\left[\mathcal{N}\left(\boldsymbol{x}_{j},\,2c_{1}^{2}\bar{P}^{\frac{1}{\alpha}}\log^{2}n\right)\right]\right\} $
approaches $0$ as $n\rightarrow\infty$. Therefore a.a.s. the number
of nodes competing with node $j$ where $j\in\mathcal{O}$ is smaller
than or equal to $\left(1+\delta_{3}\right)4\pi c_{1}^{4}\bar{P}^{\frac{2}{\alpha}}\log^{4}n$.
Let $\eta_{j}^{h}$ be the event that node $j$, $j\in\mathcal{O}$,
is active. It can be shown that for any $j\in\mathcal{O}$,
\begin{align}
 & \Pr\left[\eta_{j}^{h}\right]\nonumber \\
\geq & \int_{0}^{\infty}\left(e^{-\lambda_{l}t}\right)^{\pi\left(\sqrt{10}c_{1}\bar{P}^{\frac{1}{\alpha}}\log n\right)^{2}}\left(e^{-\lambda_{h}t}\right)^{\left(1+\delta_{3}\right)4\pi c_{1}^{4}\bar{P}^{\frac{2}{\alpha}}\log^{4}n}\nonumber \\
 & \times\lambda_{h}e^{-\lambda_{h}t}\mathrm{d}t\nonumber \\
= & \frac{1}{\pi\left(\sqrt{10}c_{1}\bar{P}^{\frac{1}{\alpha}}\log n\right)^{2}\frac{\lambda_{l}}{\lambda_{h}}+1\left(1+\delta_{3}\right)4\pi c_{1}^{4}\bar{P}^{\frac{2}{\alpha}}\log^{4}n+1}\nonumber \\
= & \frac{1}{\pi\left(\sqrt{10}c_{1}\bar{P}^{\frac{1}{\alpha}}\right)^{2}\log^{4}n+\left(1+\delta_{3}\right)4\pi c_{1}^{4}\bar{P}^{\frac{2}{\alpha}}\log^{4}n+1}\label{eq:transmission opportunities of high power links}
\end{align}

\bibliographystyle{IEEEtran}

\begin{IEEEbiography}
[{\includegraphics[width=1in,height=1.25in,clip,keepaspectratio]{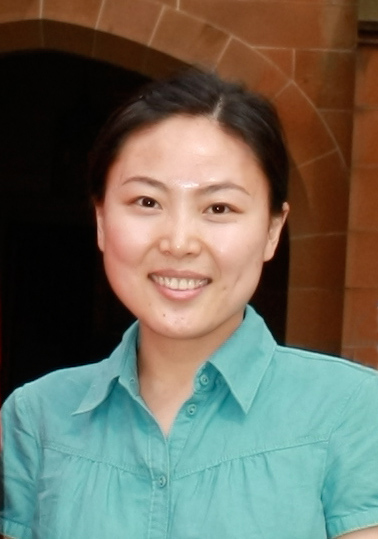}}]{Tao Yang}
received her BEng degree and MSc degree in Electrical Engineering from Southwest Jiaotong University, China. She is currently working towards the PhD degree in Engineering at The University of Sydney. Her research interests include wireless multihop networks and network performance analysis.
\end{IEEEbiography}

\begin{IEEEbiography}
[{\includegraphics[width=1in,height=1.25in,clip,keepaspectratio]{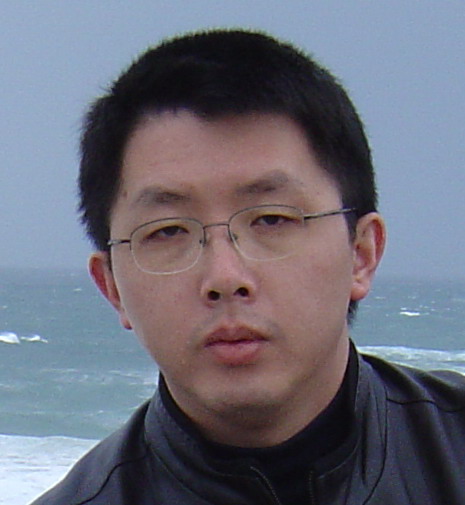}}]{Guoqiang Mao} (S'98-M'02-SM'08) received PhD in telecommunications engineering in 2002 from Edith Cowan University. Between 2002 and 2013, he was a Lecturer, a Senior Lecturer and an Associate Professor at the School of Electrical and Information Engineering, the University of Sydney, all in tenured positions. He currently holds the position of Professor of Wireless Networking, Director of Center for Real-time Information Networks at the University of Technology, Sydney. He has published more than 100 papers in international conferences and journals, which have been cited more than 2000 times. His research interest includes intelligent transport systems, applied graph theory and its applications in networking, wireless multihop networks, wireless localization techniques and network performance analysis. He is a Senior Member of IEEE, an Editor of IEEE Transactions on Vehicular Technology and IEEE Transactions on Wireless Communications and a co-chair of IEEE Intelligent Transport Systems Society Technical Committee on Communication Networks.
\end{IEEEbiography}

\begin{IEEEbiography}
[{\includegraphics[width=1in,height=1.25in,clip,keepaspectratio]{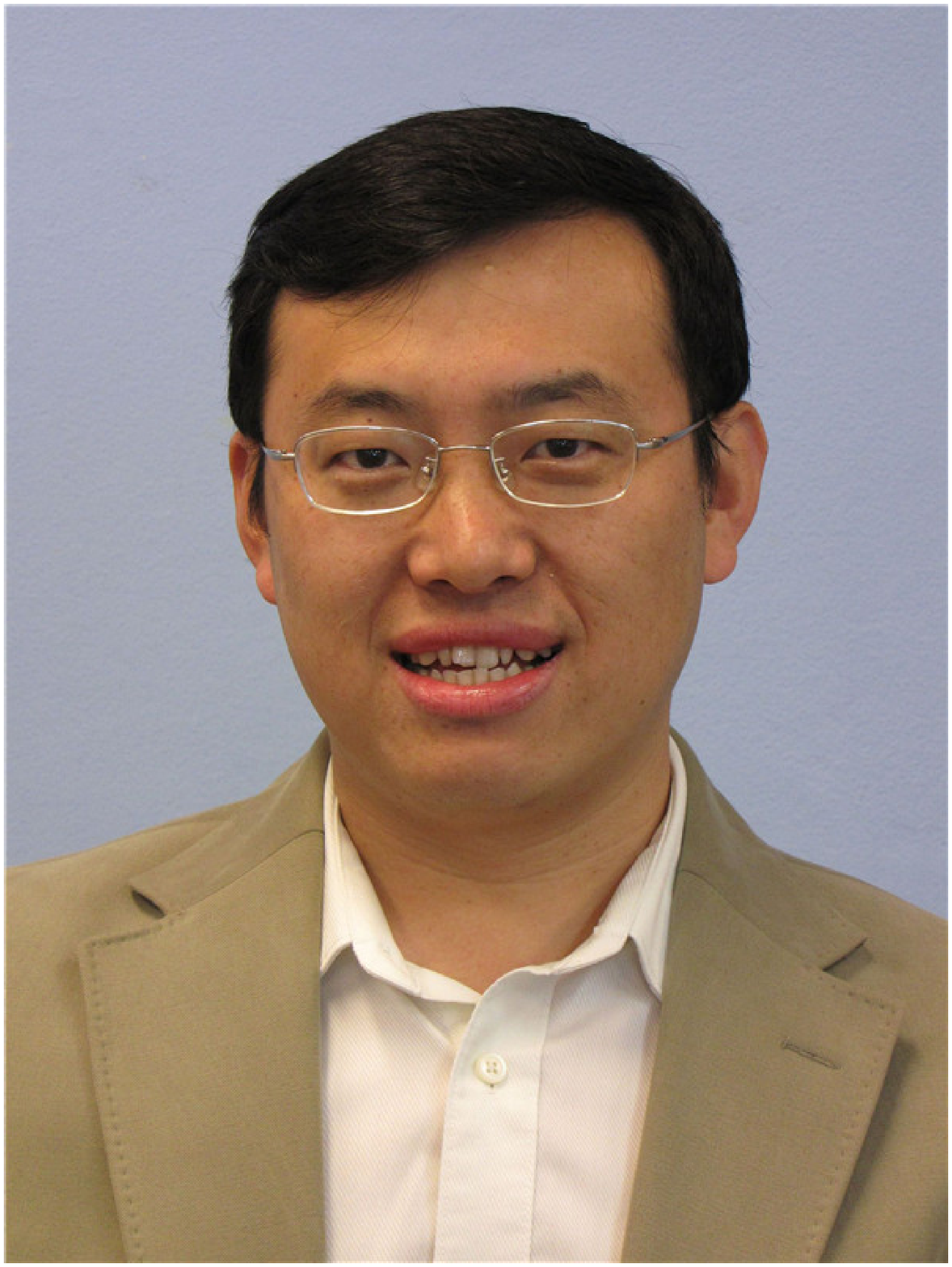}}]{Wei Zhang}
(S'01-M'06-SM'11) received the Ph.D. degree in electronic engineering from The Chinese University of Hong Kong in 2005. He was a Research Fellow with the Department of Electronic and Computer Engineering, The Hong Kong University of Science and Technology, during 2006-2007. From 2008, he has been with the School of Electrical Engineering and Telecommunications, The University of New South Wales, Sydney, Australia, where he is an Associate Professor. His current research interests include cognitive radio, cooperative communications, space-time coding, and multiuser MIMO.

He received the best paper award at the 50th IEEE Global Communications Conference (GLOBECOM), Washington DC, USA, in 2007 and the IEEE Communications Society Asia-Pacific Outstanding Young Researcher Award in 2009. He was TPC Co-Chair of Communications Theory Symposium of the IEEE International Conference on Communications (ICC), Kyoto, Japan, in 2011. He serves TPC Chair of the Symposium on Signal Processing for Cognitive Radios and Networks in the 2nd IEEE Global Conference on Signal and Information Processing (GlobalSIP) -  Atlanta, USA, in 2014. He is an Editor of the IEEE TRANSACTIONS ON WIRELESS COMMUNICATIONS and an Editor of the IEEE JOURNAL ON SELECTED AREAS IN COMMUNICATIONS (Cognitive Radio Series).
\end{IEEEbiography}

\begin{IEEEbiography}
[{\includegraphics[width=1in,height=1.25in,clip,keepaspectratio]{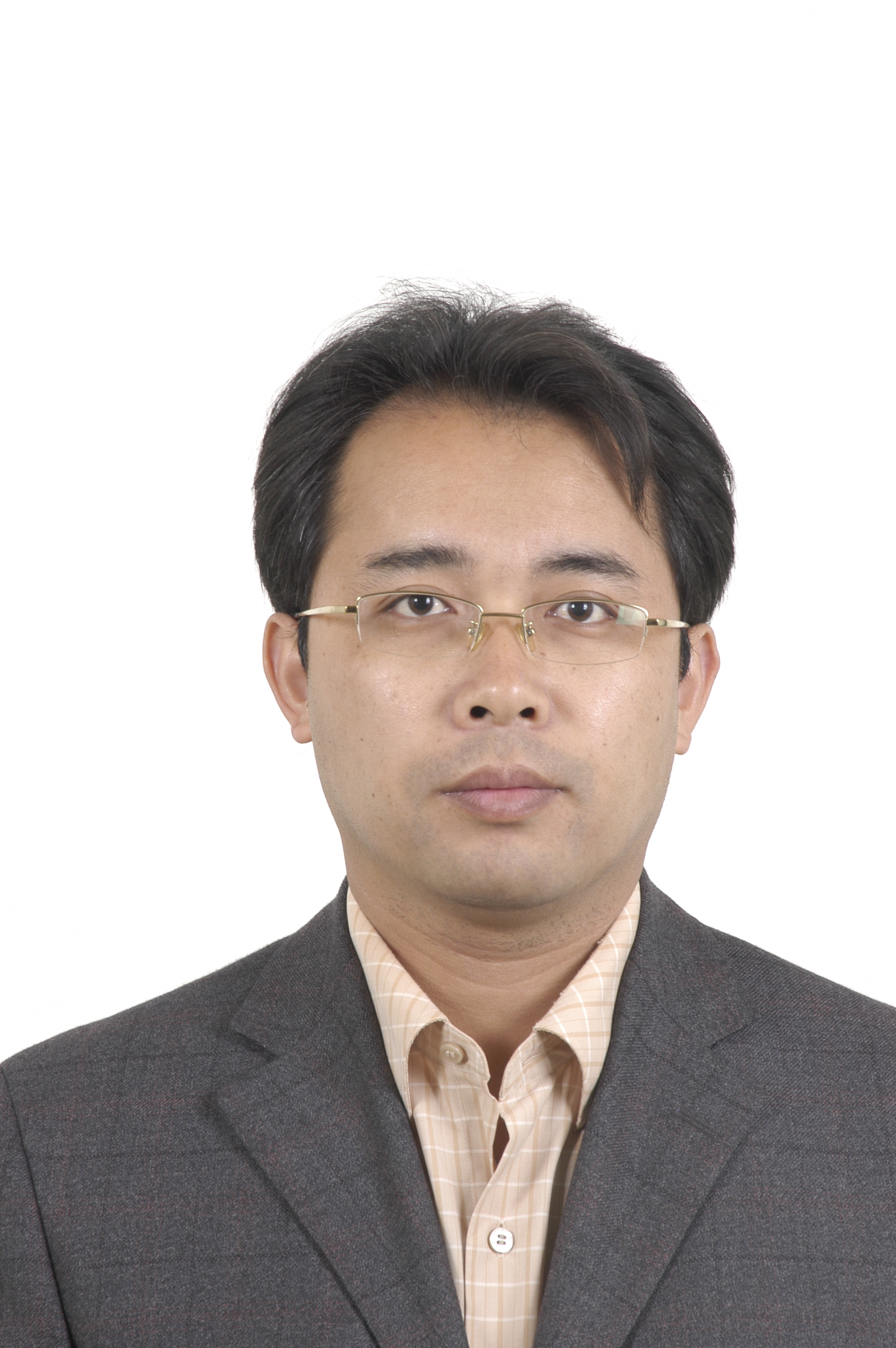}}]{Xiaofeng Tao}
 (FIET, SMIEEE) received his B.S degree in electrical engineering from Xi'an Jiaotong University, China, in 1993, and M.S.E.E. and Ph.D. degrees in telecommunication engineering from Beijing University of Posts and Telecommunications (BUPT) in 1999 and 2002, respectively. He was a visiting Professor at Stanford University from 2010 to 2011, chief architect of Chinese National FuTURE 4G TDD working group from 2003 to 2006 and established 4G TDD CoMP trial network in 2006. He is currently a Professor at BUPT, the inventor or co-inventor of 50 patents, the author or co-author of 120 papers, in 4G and beyond 4G.
\end{IEEEbiography}

\end{document}